\documentclass[reprint, amsmath,amssymb, aps, prl, superscriptaddress, nobibnotes]{revtex4-2}

\usepackage{graphicx}
\usepackage{dcolumn}
\usepackage{bm}
\usepackage{amsmath}
\usepackage{braket}
\usepackage{hyperref} 
\usepackage{subcaption}
\usepackage[utf8]{inputenc}
\usepackage{xcolor}
\usepackage{array}      %
\usepackage{multirow}   %
\usepackage{ragged2e}
\usepackage{float}
\usepackage{chngcntr}  %

\newcommand{\affa}{State Key Laboratory of Low Dimensional Quantum Physics, Department of Physics, Tsinghua University, Beijing 100084, China}
\newcommand{\affb}{Future Mobile Technology Lab, China Mobile Communications Research Institute, Beijing 100053, China}
\newcommand{\affc}{Beijing Academy of Quantum Information Sciences, Beijing 100193, China}
\newcommand{\affd}{Hefei National Laboratory, Hefei 230088, China}
\newcommand{\affe}{Frontier Science Center for Quantum Information, Beijing 100084, China}

\begin{document}

\preprint{APS/123-QED}

\title{Precision Polarization Tuning for Light Shift Mitigation in Trapped-Ion Qubits}


\author{Hengchao Tu}
\email{Corresponding author: thc20@mails.tsinghua.edu.cn}
\affiliation{\affa}
\author{Chun-Yang Luan}
\thanks{The first two authors contributed equally to this work}
\affiliation{\affb}
\author{Menglin Zou}
\affiliation{\affa}
\author{Zihan Yin}
\affiliation{\affa}
\author{Kamran Rehan}
\affiliation{\affa}
\affiliation{\affc}
\author{Kihwan Kim}
\email{Corresponding author: kimkihwan@mail.tsinghua.edu.cn}
\affiliation{\affa}
\affiliation{\affc}
\affiliation{\affd}
\affiliation{\affe}




             
\begin{abstract}
Trapped-ion qubits are among the most promising candidates for quantum computing, quantum information processing, and quantum simulation. 
In general, trapped ions are considered to have sufficiently long coherence times, which are mainly characterized under laser-free conditions. 
However, in reality, essential laser fields for quantum manipulation introduce residual light shift, which seriously degrades the coherence due to power fluctuations. Here, we present a comprehensive study of AC Stark shifts in the hyperfine energy levels of the $^{171}\mathrm{Yb}^+$ ion, revealing an asymmetric light shift between two circular polarizations in the clock qubit and pronounced vector light shifts in the Zeeman qubits. By precisely tuning these polarizations, a remarkable enhancement in coherence time is observed, reaching over a hundredfold for the clock qubit and more than tenfold for the Zeeman qubits, when comparing conditions of maximum and minimum shifts. These findings advance the practical realization of scalable trapped-ion quantum processors, enabling deep quantum circuit execution and long-duration adiabatic operations.


\end{abstract}


\maketitle
Trapped ions represent one of the most promising platforms for quantum computation and simulation, characterized by exceptional coherence times~\cite{wang2021single, wang2017single} and high-fidelity single- and two-qubit gate operations~\cite{smith2024single,clark2021high, loschnauer2024scalable}. Although such remarkable coherence has predominantly been demonstrated under laser-free conditions, practical entangling operations typically necessitate laser-mediated coupling to collective vibrational modes~\cite{leibfried2003quantum, haffner2008quantum,chen2021quantum,cai2023entangling}. This laser-induced coupling inevitably introduces residual light shifts if imperfectly compensated. These uncompensated shifts render the coherence times vulnerable to laser power fluctuations, significantly restricting the achievable circuit depth in quantum computations, limiting the feasibility of prolonged adiabatic operations in quantum simulations~\cite{haffner2003precision,blatt2012quantum, monroe2021programmable} and the long interrogation time for precision quantum metrology~\cite{gan2018oscillating}.

To mitigate these issues, two primary techniques have been developed: magic wavelength~\cite{ye1999trapping,katori1999magneto,ido2003recoil,mckeever2003state}, and magic polarization~\cite{choi2007elimination,flambaum2008magic}. 
Magic wavelength eliminates differential light shifts in optical transitions, significantly enhancing the precision and stability of optical lattice clocks~\cite{liu2015measurement, brown2017hyperpolarizability} and dipole traps~\cite{rengelink2018precision}. Similarly, magic polarization employs a specific laser polarization to cancel the differential light shift, thereby mitigating Stark-induced dephasing effects in optical trapping~\cite{kim2013magic, jackson2019magic}, and enabling precise manipulation of alkali-metal atoms~\cite{cho2023use}. 
While magic wavelength and magic polarization techniques are now widely employed in neutral atoms, their implementation and development in trapped ions have received comparatively little attention. 
With the profound progress in trapped-ion systems for quantum simulation and quantum computing, the significance of mitigating AC Stark shifts has started to gain recognition. 
Recently, measuring the magic wavelength of $^{40}\text{Ca}^{+}$ ions has been proposed as a promising strategy for developing high-precision optical clocks~\cite{liu2015measurement}. Furthermore, magic polarization has been successfully employed to protect quantum information encoded in $^{133}\text{Ba}^{+}$ ground-state qubit within the $omg$ protocol~\cite{vizvary2024eliminating}.

Clock and Zeeman states are widely used as qubits in trapped-ion quantum computing and quantum simulation. Particularly the clock qubits, insensitive to first-order Zeeman effects, have demonstrated extended coherence times for high-precision operations~\cite{wang2021single, wang2017single}. 
In this study, we investigate the AC Stark shift in the clock and Zeeman states within the $^2S_{1/2}$ manifold of the $^{171}\text{Yb}^{+}$ ion induced by various polarizations of a 355 nm picosecond pulsed laser~\cite{mizrahi2014quantum,hayes2010entanglement,campbell2010ultrafast}. We observe an unexpected asymmetric shift between $\sigma_+$ and $\sigma_-$ polarizations in the clock qubit, as well as vector shifts in the two Zeeman qubits.
Based on a detailed characterization of these polarization-dependent AC Stark shift, we identify specific polarization conditions that minimize the differential shifts, which leads to a substantial improvement in coherence time—approximately a hundredfold enhancement for the clock qubit and a tenfold enhancement for the Zeeman qubit when comparing conditions of maximum and minimum shifts.

\begin{figure}[htbp]
    \centering
    \captionsetup{justification=raggedright}
    \includegraphics[width=0.49\textwidth]{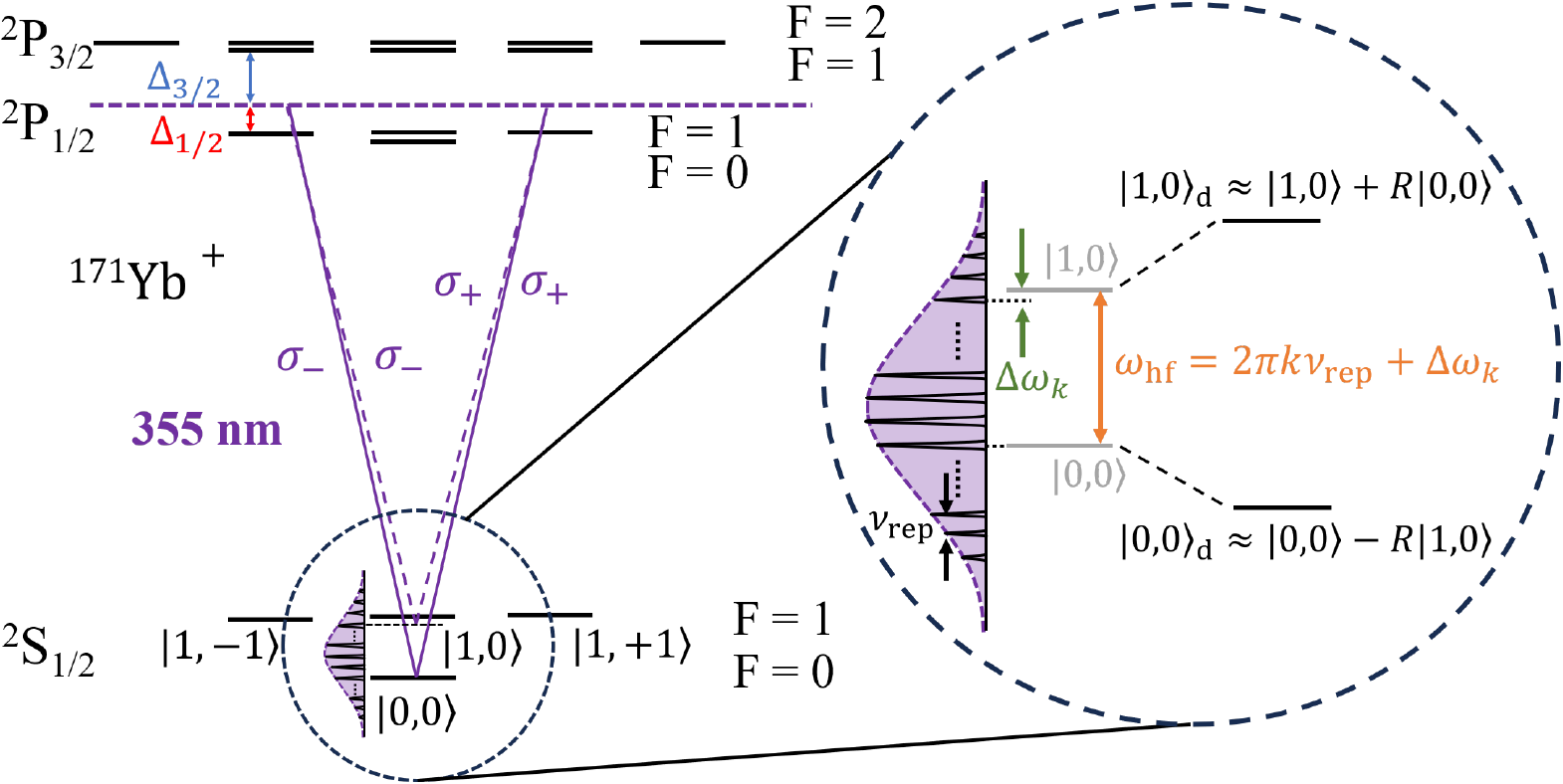}  
    \put(-250, 127){\textbf{ (a) }} 
    \put(-97, 102){\textbf{  (b) }} 
    \caption{\justifying (a) Energy level diagram of the $^{171}\text{Yb}^{+}$ ion. The 355 nm laser operates at a frequency between the $^2P_{1/2}$ and $^2P_{3/2}$ levels, thereby inducing a second-order AC Stark shift in the four hyperfine sub-levels of the $^2S_{1/2}$ ground state. This laser is pulsed with a repetition rate $\nu_\text{rep}$ in the picosecond regime, enabling spectral components that span the hyperfine splitting and drive two-photon Raman transitions. These Raman processes result in fourth-order AC Stark shifts, with detunings of $\Delta\omega_k = \omega_{\mathrm{hf}} - 2\pi k \nu_{\mathrm{rep}}$, which is described in detail in (b). In the experiment, the laser polarization is chosen to exclude any $\pi$-polarized components, thereby preventing fourth-order AC Stark shifts among the Zeeman sub-levels. (b) The clock states are modified by an external magnetic field $\mathbf{B}$, leading to the dressed state $\ket{F,m_F}_\text{d}$, which is the mixture of clock states and $R = \mu_{B}B/w_{\mathrm{hf}}$.}
    \label{fig:Yb energy level}
\end{figure}

In the $^2S_{1/2}$ ground-state manifold of the $^{171}\text{Yb}^{+}$ ion, the clock qubit is encoded in the hyperfine states $\ket{F=0, m_F=0}$ and $\ket{F=1, m_F=0}$, which are separated by a hyperfine splitting of $\omega_{\mathrm{hf}} = (2\pi)~12.642812$ GHz
in the absence of a magnetic field and Zeeman qubits, characterized by transitions with $\Delta m_F = \pm 1$, are typically defined between the states $\ket{F=1, m_F = \pm 1}$ and $\ket{F=0, m_F = 0}$, as illustrated in Fig.~\ref{fig:Yb energy level}(a). These qubits are coherently manipulated by a 355 nm picosecond pulsed laser, where the frequency resides between the $^2P_{1/2}$ and $^2P_{3/2}$ states, with the detunings of $\Delta_{1/2}=+(2\pi)~33~\mathrm{THz}$ and $\Delta_{3/2}=-(2\pi)~67~\mathrm{THz}$, respectively. We note that these detunings allow the laser to couple both states with comparable strengths. The picosecond pulsed laser generates a frequency comb with a repetition rate $\nu_{\mathrm{rep}}$, whose spectral components span the hyperfine splitting, thereby enabling the implementation of both single-qubit and multi-qubit gates via a two-photon Raman process~\cite{campbell2010ultrafast, hayes2010entanglement, wong2017demonstration, lu2019global, figgatt2019parallel}.

Our report focuses on characterizing the AC Stark shifts induced by a single 355 nm picosecond pulsed laser, which is critical for achieving high-precision frequency control in quantum gate operations while maintaining qubit coherence. For the clock qubit, second-order differential shifts, denoted as $\delta E^{(2)}_{10,00}$, arise from off-resonant coupling to excited electronic states. Additionally, the frequency comb nature of the pulsed laser gives rise to a fourth-order AC Stark shift, $\delta E^{(4)}_{10,00}$, determined by the detunings $\Delta \omega_k = w_{\mathrm{hf}} - 2\pi k \nu_{\mathrm{rep}}$, where $k$ indexes the modes of the frequency comb~\cite{lee2016engineering}, as illustrated in Fig.~\ref{fig:Yb energy level}(b). As a result, the total differential shift is the sum of both the second- and fourth-order contributions: 
\begin{align} 
\delta E^{\text{total}}_{10,00} = \delta E^{(2)}_{10,00} + \delta E^{(4)}_{10,00}. 
\label{equ:total ac Stark shift} 
\end{align}

\begin{figure}[htbp]
    \centering
    \captionsetup{justification=raggedright}
    \begin{subfigure}[b]{0.89\linewidth}
        \centering
        \includegraphics[width=\linewidth]{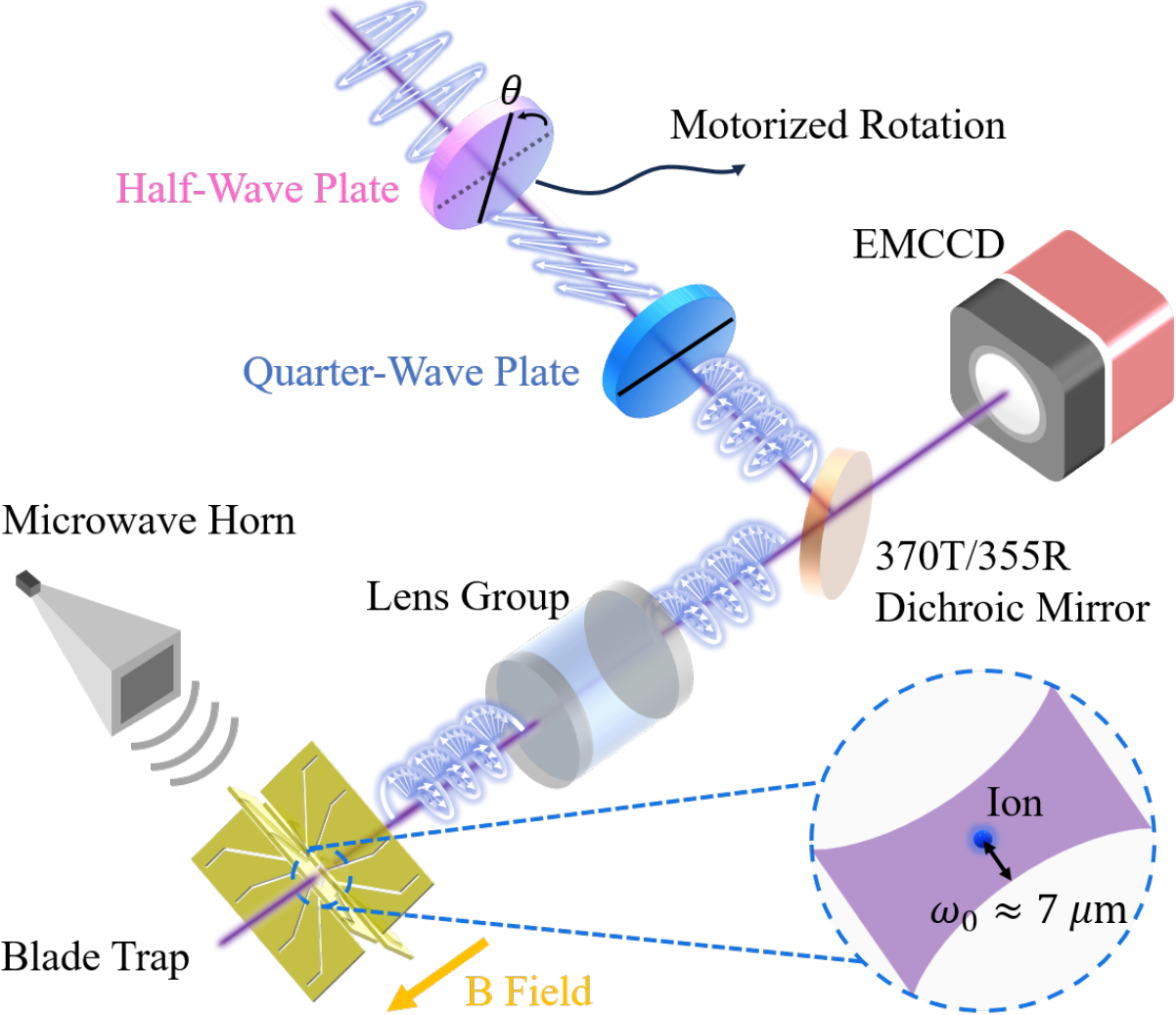}
        \put(-240, 170){\textbf{ (a) }} 
    \end{subfigure}

    \begin{subfigure}[b]{0.95\linewidth}
        \centering
        \includegraphics[width=\linewidth]{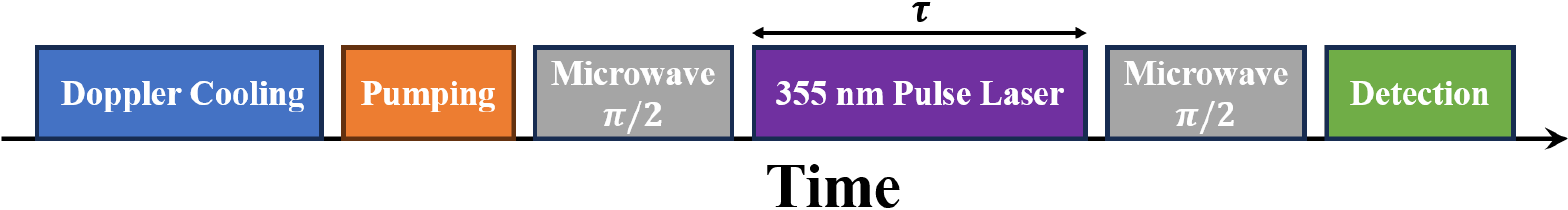} 
    \put(-240, 35){\textbf{ (b) }} 
    \end{subfigure}
    
    \caption{\justifying (a) Diagram of the optical path. HWP and QWP are used to adjust the polarization of the laser via the motorized rotation, and $\theta$ is the relative angle between the HWP and QWP. The 355 nm laser is focused onto the ion with a waist radius ($\omega_{0}$) of approximately $7~\mathrm{\mu m}$. (b) The Ramsey measurement sequence for the AC Stark shift measurement is realized by scanning the time interval $\tau$ between two $\pi$/2 microwave pulses. }  
    \label{fig:system and sequence}
\end{figure}

An external magnetic field $\mathbf{B}$ is typically applied to define the quantization axis. In our setup, $\mathbf{B}$ is aligned along the propagation direction of the 355 nm pulsed laser, thereby suppressing $\pi$-polarized transitions as shown in Fig.~\ref{fig:system and sequence}(a). The bare clock states are modified into dressed states with $\mathbf{B}$~\cite{derevianko2010theory}, as written by: 
\begin{align}
    \ket{1,0}_\text{d} &\approx \ket{1, 0} + \frac{\mu_{B}B}{w_{\mathrm{hf}}} \ket{0, 0}  \notag  \\
    \ket{0,0}_\text{d} &\approx \ket{0, 0} - \frac{\mu_{B}B}{w_{\mathrm{hf}}} \ket{1, 0},
\label{equ:dress state}
\end{align}
where $\ket{F,m_F}$ denotes the bare states and $\ket{F,m_F}_\text{d}$ represents the corresponding dressed states as illustrated in Fig.~\ref{fig:Yb energy level}(b). Magnetic-field-induced hyperfine mixing modifies the second-order differential shift(see Supplementary materials~\cite{Supp} for further details), which can be expressed as~\cite{wineland2003quantum,campbell2010ultrafast,lee2016engineering}: 
\begin{flalign}
\delta E_{\text{10,00}}^{(2)} &\equiv \Delta E_{10}^{(2)} -  \Delta E_{00}^{(2)} \notag\\ 
&= \left( \epsilon_+^2 +\epsilon_-^2 \right)\frac{\omega_{\mathrm{hf}}}{12} \left( \frac{g^2_{1/2}}{\Delta_{1/2}^2} + \frac{2g^2_{3/2}}{\Delta_{3/2}^2} \right)I \notag \\
& \quad - \left( \epsilon_+^2 -\epsilon_-^2 \right) \frac{4\mu_B B}{12\omega_{\mathrm{hf}}}  \left ( \frac{g^2_{1/2}}{\Delta_{1/2} } - \frac{g^2_{3/2}}{\Delta_{3/2}}  \right)I,  \notag \\ 
\label{equ:clock qubit 2nd order}
\end{flalign}
where $\Delta E_{10}^{(2)}$ and $\Delta E_{00}^{(2)}$ are the second-order AC Stark shifts for the $\ket{1,0}$ and $\ket{0,0}$ states~\cite{Supp}, respectively. The laser polarization vector is parameterized as $\hat{\epsilon} = \epsilon_+ \hat{\sigma}_+ + \epsilon_- \hat{\sigma}_- + \epsilon_0 \hat{\pi}$, subject to the normalization condition $\left| \epsilon_+ \right|^2 + \left| \epsilon_- \right|^2 + \left| \epsilon_0 \right|^2 = 1$. Due to the laser propagating parallel to $\mathbf{B}$, the $\pi$-polarized component vanishes, i.e., $\epsilon_0 = 0$. Here, $g_{J}^2$ ($J=1/2,3/2$) quantifies the coupling strengths to the $^2P_{J}$ excited states, $\mu_B$ is the Bohr magneton, $I = P/(\pi \omega_0^2)$ denotes the laser intensity, and $\omega_0$ is the beam waist.
This result demonstrates that $\delta E_{\text{10,00}}^{(2)}$ exhibits a polarization-dependent light shift that is proportional to the magnetic field strength $B$.

The fourth-order AC Stark shift arises from multiple detuning components $\Delta \omega_k = \omega_{\mathrm{hf}} - 2\pi k \nu_{\mathrm{rep}}$ due to the presence of a frequency comb. The resulting fourth-order differential shift can be expressed as~\cite{lee2016engineering, mizrahi2014quantum}: 
\begin{align}
\delta E_{10,00}^{(4)} &=  \left( \epsilon_+^2 - \epsilon_-^2\right)^2\frac{\mathbf{\mathcal{C}}_{10,00}}{72\Delta \omega _{\mathrm{min}}} \left(\frac{g_{1/2}^2}{\Delta_{1/2}} - \frac{g_{3/2}^2}{\Delta_{3/2}}  \right)^2 I^2.
\label{equ:clock qubit 4th order}
\end{align}
Here $\Delta \omega_{\mathrm{min}}$ is the minimum detuning, defined as $\Delta \omega_{\mathrm{min}} = \text{min}\left| \omega_{\mathrm{hf}} - 2\pi k \nu_{\mathrm{rep}} \right| = \omega_{\mathrm{hf}} - 2\pi n \nu_{\mathrm{rep}}$, with $n$ denoting the index corresponding to the nearest comb mode. And the frequency-comb factor $\mathcal{C}_{10,00}$ is given by: 
\begin{align}
\mathbf{\mathcal{C}}_{00,10} &= \sum^{\infty}_{k=-\infty} \frac{ \text{sech}^2 \left( (n+k)\pi \nu_{\mathrm{rep}}\tau_\mathrm{p}\right) }{1 - \frac{2\pi k\nu_{\mathrm{rep}}}{\Delta\omega_\mathrm{min}}},
\label{equ:raman process coefficient}
\end{align}
where $\tau_p$ denotes the pulse duration. 
In the fourth-order AC Stark shift, the influence of dressed-state corrections becomes the second order of $\mu_B B/\omega_{\mathrm{hf}}$, which results in a negligible amount. In short, the total differential shift of the clock qubit consists of both second- and fourth-order contributions, which are significant under different laser polarizations.

For Zeeman qubits, the total differential shift is dominated by the second-order contribution. Since no $\pi$-polarized component is present in our configuration, the fourth-order AC Stark shift vanishes. Therefore, the total shift can be written as~\cite{Supp}:
\begin{flalign}
\delta E_{\text{1±1,00}}^{\text{total}} &\equiv \Delta E_{\text{1±1}}^{(2)} -  \Delta E_{00}^{(2)} \notag\\ 
&= \pm\left( \epsilon_-^2 -\epsilon_+^2 \right)\frac{1}{12}\left ( \frac{g^2_{1/2}}{\Delta_{1/2} } - \frac{g^2_{3/2}}{\Delta_{3/2}}  \right)I,  \notag \\ 
\label{equ:Zeeman qubits ac Stark shift}
\end{flalign} 
where $\Delta E_{1\pm1}$ denote the second-order AC Stark shifts for the $\ket{1,\pm1}$ states. The terms of $\Delta E_{1\pm1}$ is usually written in the spherical‑tensor formalism~\cite{manakov1986atoms,flambaum2008magic,le2013dynamical}: 
\begin{align}
\Delta E_{1m_F}^{(2)}  &= \frac14 |E|^2 \Bigg[
\alpha_{nJF}^s + (\epsilon_-^2 - \epsilon_+^2)\,\alpha_{nJF}^v \frac{m_F}{F} \notag\\
 &+ (3\epsilon_\pi^2-1)\,\alpha_{nJF}^T\frac{3m_F^2 - F(F+1)}{2F(2F-1)}\Bigg],
\label{equ:Zeeman state dynamical polarizability}
\end{align}
where $\alpha_{nJF}^{s}$, $\alpha_{nJF}^{v}$, and $\alpha_{nJF}^{T}$ are the scalar, vector, and tensor polarizability, respectively. $E$ is the electric‑field amplitude, and $m_F=\pm1$. For differential shifts $\delta E_{\text{1±1,00}}^{\text{total}}$, the scalar polarizability $\alpha_{nJF}^{s}$ is negligible, where the $\Delta_{1/2}$ and $\Delta_{3/2}$ is much larger than hyperfine splitting $\omega_{\mathrm{hf}}$ as for 355 nm laser with the $^{171}\text{Yb}^{+}$ ion. The tensor polarizability $\alpha_{nJF}^{T}$ is zero for $^2S_{1/2}$ manifold of the $^{171}\text{Yb}^{+}$ ion~\cite{le2013dynamical}. Therefore, the term of vector polarizability $\alpha_{nJF}^{v}$, which shows clearly polarization dependence, is in agreement with Eq.~(\ref{equ:Zeeman qubits ac Stark shift}). 

We experimentally study the the polarization-dependent AC Stark shift of a clock qubit and two Zeeman qubits of trapped $^{171}\text{Yb}^{+}$ ion, induced by 355 nm laser with a pulse duration $\tau_\mathrm{p}$ of 12.883 ps and a repetition rate $\nu_{\mathrm{rep}}$ of 118.993 MHz. As shown in Fig.~\ref{fig:system and sequence}(a), the polarization can be controlled using a quarter-wave plate (QWP) and a half-wave plate (HWP). The external magnetic field has a magnitude of 11.343~Gauss, oriented parallel to the propagation direction of the laser. The AC Stark shift is measured using the microwave Ramsey sequence, as illustrated in Fig.~\ref{fig:system and sequence}(b).
\begin{figure}[htbp]
    \centering
    \captionsetup{justification=raggedright}
    \includegraphics[width=0.48\textwidth]{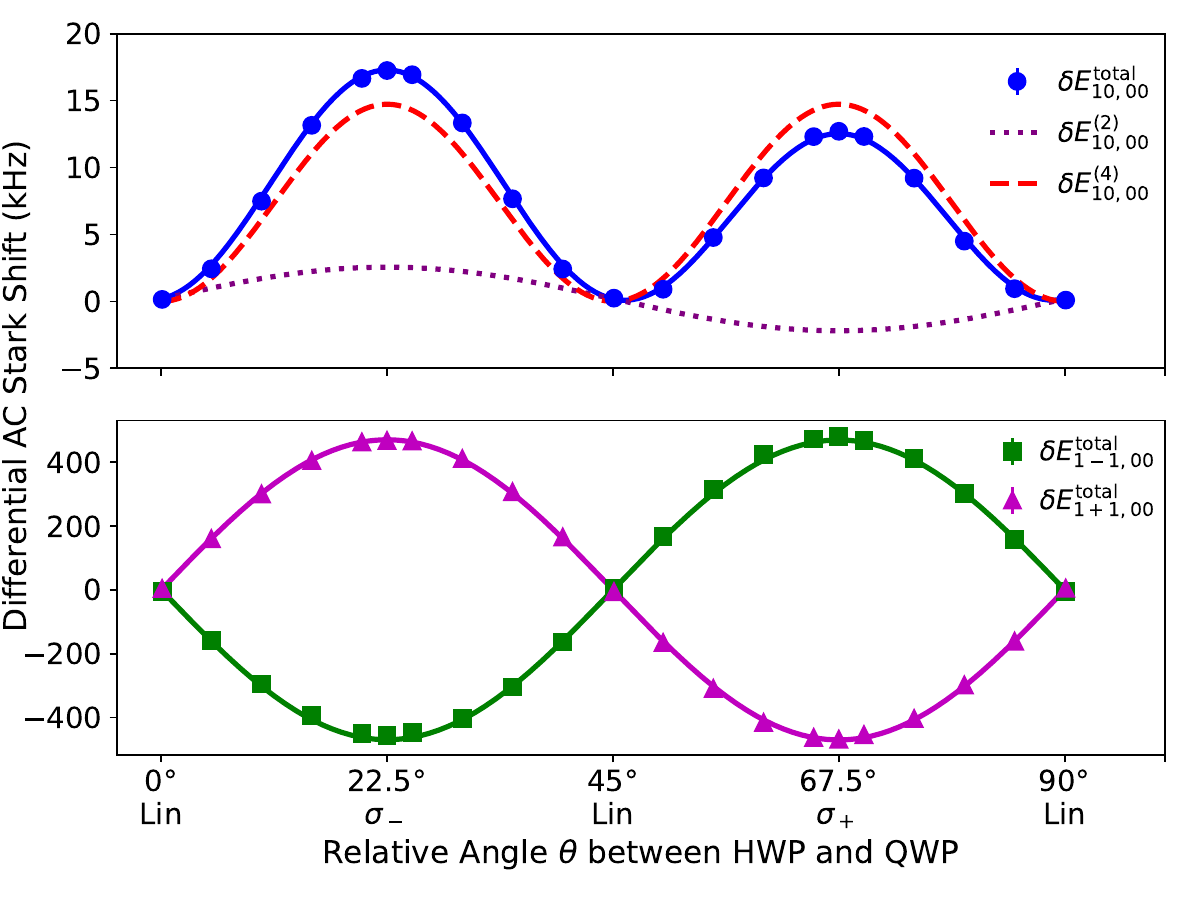}  
    \put(-220, 164){\textbf{ (a) }}
    \put(-220,  83){\textbf{ (b) }}
    \caption{\justifying Differential AC Stark shift as a function of relative angle $\theta$ between the HWP and QWP. (a) The clock qubit $\Delta m_F=0$ (filled circles) exhibits a pronounced asymmetry in AC Stark shifts between $\sigma_+$ and $\sigma_-$ polarizations, with the shift induced by $\sigma_-$ being larger than that of $\sigma_+$. The solid, dotted, and dashed lines represent total, second- and fourth-order differential shifts, respectively, calculated using Eqs.~\eqref{equ:total ac Stark shift}, \eqref{equ:clock qubit 2nd order} and \eqref{equ:clock qubit 4th order}. (b) The Zeeman qubit $\Delta m_F=+1$ (triangles) and $\Delta m_F=-1$ (squares) exhibit shifts over an order of magnitude larger than those observed in the clock states. These shifts demonstrate a clear vector‑Stark‑shift behavior: $\Delta m_F=+1$ ($-1$) states experience positive (negative) shifts under $\sigma_-$ polarization and negative (positive) shifts under $\sigma_+$ polarization. The solid curves show theoretical predictions from Eq.~\eqref{equ:Zeeman qubits ac Stark shift}, with beam width $\omega_0$ fitted to experimental data. Each data point was extracted via Ramsey interferometry. Error bars (standard error) were calculated from 2000 bootstrap resamples, which are smaller than the marker size and thus not visible.}
    \label{fig:ac shift by polarization}
\end{figure}

The measured differential shifts depending on the polarization of the laser are presented in Fig.~\ref{fig:ac shift by polarization}. The polarization is tuned continuously from linear, through circular, and back to linear by rotating the HWP while keeping the QWP fixed, as illustrated in Fig.~\ref{fig:system and sequence}(a). All data are acquired with a 355 nm beam of fixed power 52.1 mW and a waist radius $\omega_0=7~\mathrm{\mu m}$, which is obtained from a fit to experimental results. For the clock qubit shown in Fig.~\ref{fig:ac shift by polarization}(a), the measured shifts (filled circles) exhibit a clear asymmetry between $\sigma_-$ and $\sigma_+$ polarizations, where $\sigma_-$ produces a substantially larger shift. The solid, dotted, and dashed are mathematical curves based on Eqs.~\eqref{equ:total ac Stark shift}, \eqref{equ:clock qubit 2nd order} and \eqref{equ:clock qubit 4th order}, respectively. This asymmetry arises from the magnetic‑field–induced second‑order AC Stark shift in Eq.~\eqref{equ:clock qubit 2nd order}. Consequently, the total differential shift cannot be minimized simply by employing an equal superposition of $\sigma_-$ and $\sigma_+$, which is linear polarization~\cite{vizvary2024eliminating}. For the Zeeman qubits ($\Delta m_F = \pm1$), shown in Fig.~\ref{fig:ac shift by polarization}(b), the data display the characteristic vector‑Stark‑shift behavior, which are in good agreement with theoretical expectation of Eq.~\eqref{equ:Zeeman qubits ac Stark shift}, where two qubits experience equal‑magnitude shifts of opposite sign for a given polarization. Moreover, their maximum shifts are nearly an order of magnitude larger than those of the clock qubit.

\begin{figure}[htbp]
    \centering
    \captionsetup{justification=raggedright}
    \includegraphics[width=0.48\textwidth]{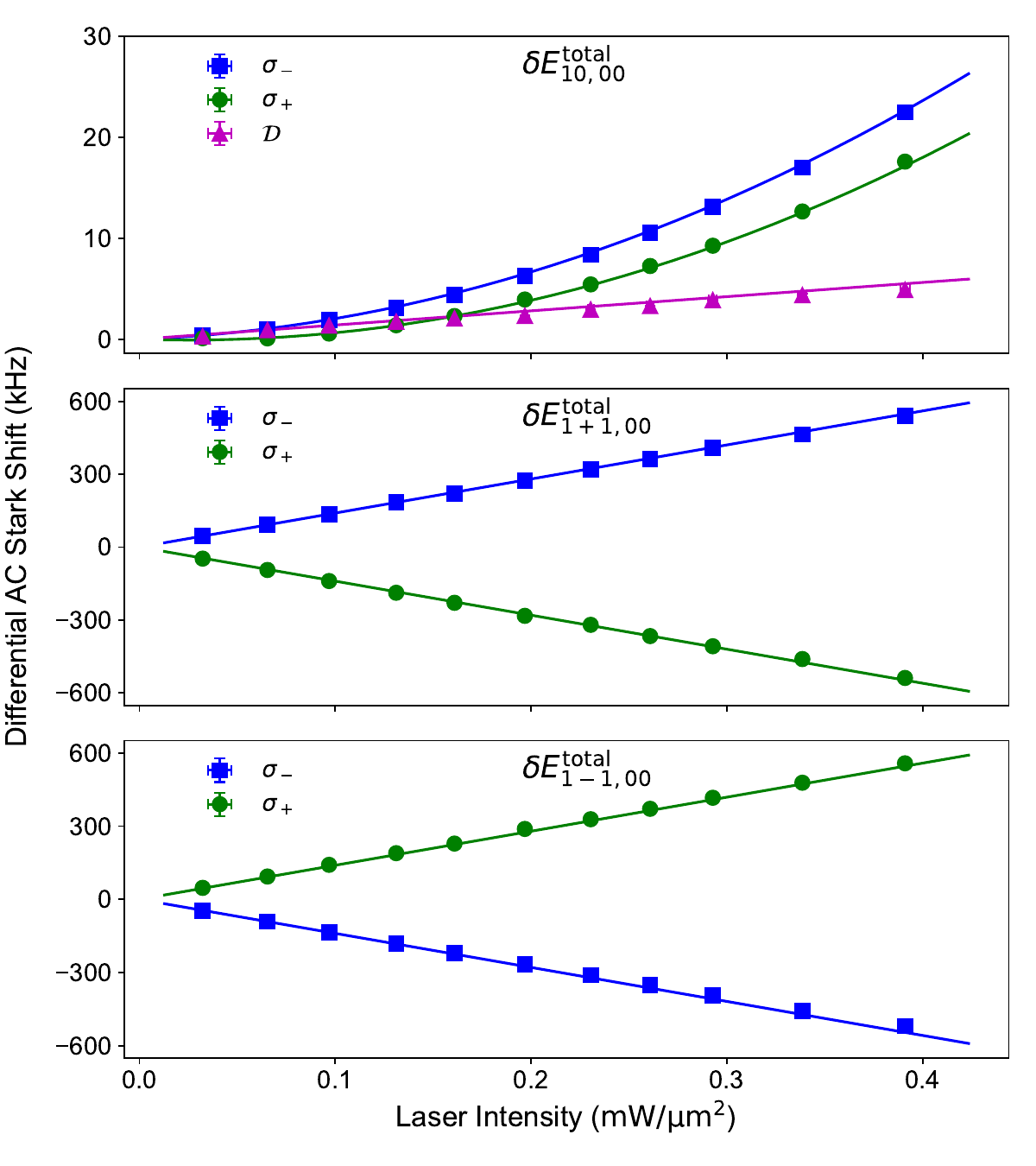}  
    \put(-217, 256){\textbf{ (a) }}
    \put(-217, 170){\textbf{ (b) }}
    \put(-217,  84){\textbf{ (c) }}
    \caption{\justifying Differential shift as a function of laser intensity. (a) The clock qubit exhibits polarization-dependent behavior in its intensity-dependent AC Stark shift. For both $\sigma_-$ (blue) and $\sigma_+$ (green), the shift is primarily determined by fourth-order light shifts, resulting in a characteristic quartic dependence on intensity. The $\mathcal{D}$(pink), defined as the shift difference between the two polarizations, displays a linear dependence on intensity, originating from the magnetic-field-dependent term in Eq.~\eqref{equ:clock qubit 2nd order}. (b) The Zeeman qubits $\Delta m_F = +1$ and (c) $\Delta m_F = -1$ exhibit a linear dependence of the AC Stark shift on laser intensity under both $\sigma_-$ (blue) and $\sigma_+$(green) polarizations, consistent with second-order light shift behavior in Eq.~\eqref{equ:Zeeman qubits ac Stark shift}. Solid lines represent theoretical predictions using beam size parameters obtained from the fit in Fig.~3. Squares, filled circles, and triangles denote experimental data points. Error bars (standard error) were calculated from 2000 bootstrap resamples, and most error bars are smaller than the marker size and thus not visible.}
    \label{fig:ac shift by power}
\end{figure}

To identify the characteristics of differential shifts, we measure the intensity dependence of light shift for $\sigma_-$($\theta$ = 22.50°) and $\sigma_+$($\theta$ = 67.50°) polarizations. For the clock qubit, where fourth‑order terms dominate, the shifts display the expected quadratic scaling with intensity as shown in Fig.~\ref{fig:ac shift by power}(a). When subtracting the differential shifts of $\sigma_-$ and $\sigma_+$ polarizations, denoted as $\mathcal{D}$, we observe a linear growth of $\mathcal{D}$ with intensity. Since it equals twice the second‑order contribution predicted by Eq.~\eqref{equ:clock qubit 2nd order}, it is direct evidence that the residual asymmetry originates from the second‑order AC Stark shift. In contrast, the Zeeman qubits are governed entirely by second‑order effects, yielding shifts that vary linearly with intensity, as shown in Fig.~\ref{fig:ac shift by power}(b,c) and in full agreement with Eq.~\eqref{equ:Zeeman qubits ac Stark shift}.

\begin{figure*}[htbp]
  \centering
  \begin{subfigure}[t]{0.49\textwidth}
    \centering
    \includegraphics[width=\linewidth]{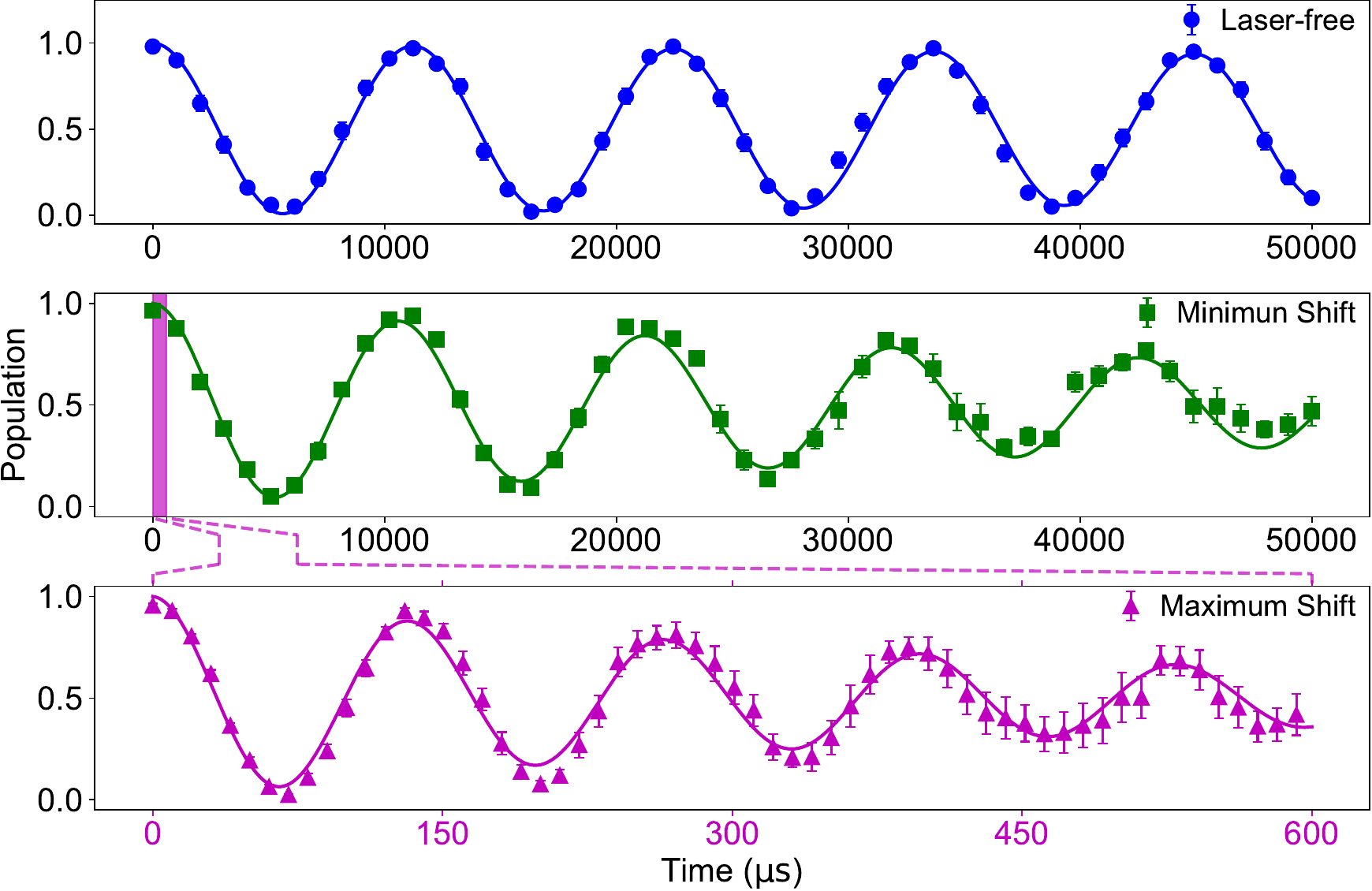}
    \put(-250, 168){\textbf{ (a) }} 
  \end{subfigure}
  \begin{subfigure}[t]{0.49\textwidth}
    \centering
    \includegraphics[width=\linewidth]{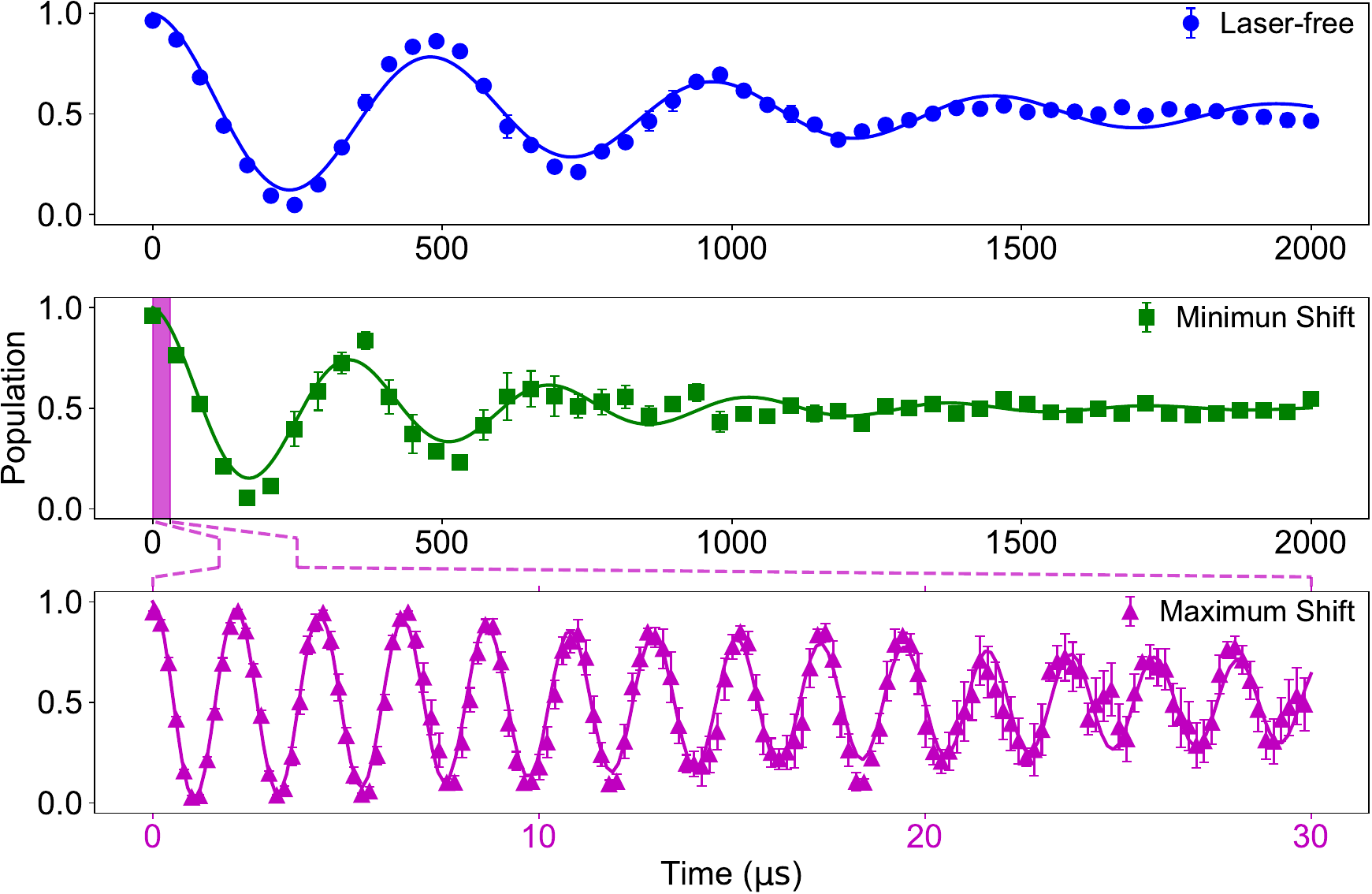}
    \put(-250, 168){\textbf{ (b) }} 
  \end{subfigure}
  \caption{\justifying (a) Ramsey measurement of clock qubit for the cases of 355 nm laser-free (top), the minimum AC Stark shift (middle), and the maximum AC Stark shift (bottom), where the corresponding coherence times are 331 ms, 55.6 ms, and 0.478 ms, respectively. The minimum and maximum AC Stark shifts are achieved by setting the $\theta$ to 45.57° and 22.40°, respectively. 
  (b) Ramsey measurement of Zeeman qubit for the cases of 355 nm laser-free (top), the minimum AC Stark shift (middle), and the maximum AC Stark shift (bottom), where the corresponding coherence times are 853~$\mathrm{\mu s}$, 467~$\mathrm{\mu s}$, and 32.6~$\mathrm{\mu s}$, respectively. The minimum and maximum AC Stark shifts are achieved by setting the $\theta$ to 45.03° and 22.50°, respectively. Solid lines represent fitted curves. Filled circles denote experimental data points. Each data point is the mean of 5 independent sets of 100 realizations (total $N = 500$); error bars denote standard error across sets.}  
  \label{fig:coherence time} 
\end{figure*}

To quantify the influence of polarization control on qubit coherence, we measure the coherence times of the clock qubit ($\Delta m_F = 0$) and the Zeeman qubit ($\Delta m_F = +1$) under three conditions: (i) 355 nm laser-free, (ii) the minimum shift, and (iii) the maximum shift. As shown in Fig.~\ref{fig:coherence time}(a), the clock-qubit coherence times are 331 ms without the laser, 55.6 ms at the minimum shift ($\theta$ = 45.57°), and 0.478 ms at maximum shift ($\theta$ = 22.40°). We emphasize that more than two orders of magnitude differences are observed depending on the minimum and the maximum shifts. In our experiment, the minimum shift exhibits a reduced coherence time compared to the ``laser-free'' case, because the configuration that yields the minimum shift still results in a residual light shift. Complete elimination of the residual shift would require a stronger magnetic field. Moreover, as the laser power increases, a correspondingly higher magnetic field strength is needed~\cite{vizvary2024eliminating}
(see Supplementary materials~\cite{Supp} for further details).

The Zeeman qubit results are shown in Fig.~\ref{fig:coherence time}(b). The coherence time without the laser is measured to be 853~$\mathrm{\mu s}$, decreasing to 467~$\mathrm{\mu s}$ at the minimal shift point ($\theta$ = 45.03°) and to 32.6~$\mathrm{\mu s}$ at the maximum shift point ($\theta$ = 22.50°), demonstrating more than a tenfold enhancement in coherence with optimal polarization. The ``laser-free" coherence time remains longer than that at the minimal shift point, as the Zeeman qubit is highly sensitive to polarization imperfections, which degrade coherence. We also note that the $\theta$ for the minimum differential shifts of the clock ($\theta$ = 45.57°) and Zeeman ($\theta$ = 45.03°) qubits are different, which indicates that the optimal polarization for minimizing Stark shifts in the clock qubit is not purely linear. 


In conclusion, we have experimentally characterized the polarization- and intensity-dependent AC Stark shifts for both clock and Zeeman qubits in a 355 nm laser-driven $^{171}$Yb$^+$ ion system. Our results reveal polarization asymmetries arising from magnetic-field-induced hyperfine state mixing and demonstrate that the optimal polarization condition deviates from the purely linear case previously assumed. This precise polarization tuning significantly extends the qubit coherence time, approaching that of the ``laser-free". These findings not only deepen the understanding of light shifts in $^{171}$Yb$^+$ ion systems but also offer practical guidance for mitigating decoherence in high-precision quantum control of other ion species. Beyond trapped-ion platforms, the insights gained here can be applied to neutral atom systems, where optical trapping and control similarly induce polarization-sensitive Stark shifts. Our results underscore the importance of polarization engineering for extending qubit coherence, enabling deeper quantum circuits and longer adiabatic evolutions essential for scalable quantum simulation and computation.

\begin{acknowledgments}
The authors acknowledge Wesley C. Campbell, Donghyun Cho, and Thomas Dellaert for valuable feedback and suggestions. This work was supported by the Innovation Program for Quantum Science and Technology under Grants No.2021ZD0301602 and the National Natural Science Foundation of China under Grants No.92065205, No.11974200, and No.62335013.

\end{acknowledgments}

\nocite{*}

\bibliography{apssamp}

\providecommand{\noopsort}[1]{}\providecommand{\singleletter}[1]{#1}%
\begin{thebibliography}{40}%
\makeatletter
\providecommand \@ifxundefined [1]{%
 \@ifx{#1\undefined}
}%
\providecommand \@ifnum [1]{%
 \ifnum #1\expandafter \@firstoftwo
 \else \expandafter \@secondoftwo
 \fi
}%
\providecommand \@ifx [1]{%
 \ifx #1\expandafter \@firstoftwo
 \else \expandafter \@secondoftwo
 \fi
}%
\providecommand \natexlab [1]{#1}%
\providecommand \enquote  [1]{``#1''}%
\providecommand \bibnamefont  [1]{#1}%
\providecommand \bibfnamefont [1]{#1}%
\providecommand \citenamefont [1]{#1}%
\providecommand \href@noop [0]{\@secondoftwo}%
\providecommand \href [0]{\begingroup \@sanitize@url \@href}%
\providecommand \@href[1]{\@@startlink{#1}\@@href}%
\providecommand \@@href[1]{\endgroup#1\@@endlink}%
\providecommand \@sanitize@url [0]{\catcode `\\12\catcode `\$12\catcode `\&12\catcode `\#12\catcode `\^12\catcode `\_12\catcode `\%12\relax}%
\providecommand \@@startlink[1]{}%
\providecommand \@@endlink[0]{}%
\providecommand \url  [0]{\begingroup\@sanitize@url \@url }%
\providecommand \@url [1]{\endgroup\@href {#1}{\urlprefix }}%
\providecommand \urlprefix  [0]{URL }%
\providecommand \Eprint [0]{\href }%
\providecommand \doibase [0]{https://doi.org/}%
\providecommand \selectlanguage [0]{\@gobble}%
\providecommand \bibinfo  [0]{\@secondoftwo}%
\providecommand \bibfield  [0]{\@secondoftwo}%
\providecommand \translation [1]{[#1]}%
\providecommand \BibitemOpen [0]{}%
\providecommand \bibitemStop [0]{}%
\providecommand \bibitemNoStop [0]{.\EOS\space}%
\providecommand \EOS [0]{\spacefactor3000\relax}%
\providecommand \BibitemShut  [1]{\csname bibitem#1\endcsname}%
\let\auto@bib@innerbib\@empty
\bibitem [{\citenamefont {Wang}\ \emph {et~al.}(2021)\citenamefont {Wang}, \citenamefont {Luan}, \citenamefont {Qiao}, \citenamefont {Um}, \citenamefont {Zhang}, \citenamefont {Wang}, \citenamefont {Yuan}, \citenamefont {Gu}, \citenamefont {Zhang},\ and\ \citenamefont {Kim}}]{wang2021single}%
  \BibitemOpen
  \bibfield  {author} {\bibinfo {author} {\bibfnamefont {P.}~\bibnamefont {Wang}}, \bibinfo {author} {\bibfnamefont {C.-Y.}\ \bibnamefont {Luan}}, \bibinfo {author} {\bibfnamefont {M.}~\bibnamefont {Qiao}}, \bibinfo {author} {\bibfnamefont {M.}~\bibnamefont {Um}}, \bibinfo {author} {\bibfnamefont {J.}~\bibnamefont {Zhang}}, \bibinfo {author} {\bibfnamefont {Y.}~\bibnamefont {Wang}}, \bibinfo {author} {\bibfnamefont {X.}~\bibnamefont {Yuan}}, \bibinfo {author} {\bibfnamefont {M.}~\bibnamefont {Gu}}, \bibinfo {author} {\bibfnamefont {J.}~\bibnamefont {Zhang}},\ and\ \bibinfo {author} {\bibfnamefont {K.}~\bibnamefont {Kim}},\ }\bibfield  {title} {\bibinfo {title} {Single ion qubit with estimated coherence time exceeding one hour},\ }\href {https://doi.org/10.1038/s41467-020-20330-w} {\bibfield  {journal} {\bibinfo  {journal} {Nat. Commun.}\ }\textbf {\bibinfo {volume} {12}},\ \bibinfo {pages} {233} (\bibinfo {year} {2021})}\BibitemShut {NoStop}%
\bibitem [{\citenamefont {Wang}\ \emph {et~al.}(2017)\citenamefont {Wang}, \citenamefont {Um}, \citenamefont {Zhang}, \citenamefont {An}, \citenamefont {Lyu}, \citenamefont {Zhang}, \citenamefont {Duan}, \citenamefont {Yum},\ and\ \citenamefont {Kim}}]{wang2017single}%
  \BibitemOpen
  \bibfield  {author} {\bibinfo {author} {\bibfnamefont {Y.}~\bibnamefont {Wang}}, \bibinfo {author} {\bibfnamefont {M.}~\bibnamefont {Um}}, \bibinfo {author} {\bibfnamefont {J.}~\bibnamefont {Zhang}}, \bibinfo {author} {\bibfnamefont {S.}~\bibnamefont {An}}, \bibinfo {author} {\bibfnamefont {M.}~\bibnamefont {Lyu}}, \bibinfo {author} {\bibfnamefont {J.-N.}\ \bibnamefont {Zhang}}, \bibinfo {author} {\bibfnamefont {L.-M.}\ \bibnamefont {Duan}}, \bibinfo {author} {\bibfnamefont {D.}~\bibnamefont {Yum}},\ and\ \bibinfo {author} {\bibfnamefont {K.}~\bibnamefont {Kim}},\ }\bibfield  {title} {\bibinfo {title} {Single-qubit quantum memory exceeding ten-minute coherence time},\ }\href {https://doi.org/10.1038/s41566-017-0007-1} {\bibfield  {journal} {\bibinfo  {journal} {Nat. Photon.}\ }\textbf {\bibinfo {volume} {11}},\ \bibinfo {pages} {646} (\bibinfo {year} {2017})}\BibitemShut {NoStop}%
\bibitem [{\citenamefont {Smith}\ \emph {et~al.}(2024)\citenamefont {Smith}, \citenamefont {Leu}, \citenamefont {Miyanishi}, \citenamefont {Gely},\ and\ \citenamefont {Lucas}}]{smith2024single}%
  \BibitemOpen
  \bibfield  {author} {\bibinfo {author} {\bibfnamefont {M.~C.}\ \bibnamefont {Smith}}, \bibinfo {author} {\bibfnamefont {A.~D.}\ \bibnamefont {Leu}}, \bibinfo {author} {\bibfnamefont {K.}~\bibnamefont {Miyanishi}}, \bibinfo {author} {\bibfnamefont {M.~F.}\ \bibnamefont {Gely}},\ and\ \bibinfo {author} {\bibfnamefont {D.~M.}\ \bibnamefont {Lucas}},\ }\bibfield  {title} {\bibinfo {title} {Single-qubit gates with errors at the $10^{-7}$ level},\ }\href {https://arxiv.org/abs/2412.04421} {\bibfield  {journal} {\bibinfo  {journal} {arXiv preprint arXiv:2412.04421}\ } (\bibinfo {year} {2024})}\BibitemShut {NoStop}%
\bibitem [{\citenamefont {Clark}\ \emph {et~al.}(2021)\citenamefont {Clark}, \citenamefont {Tinkey}, \citenamefont {Sawyer}, \citenamefont {Meier}, \citenamefont {Burkhardt}, \citenamefont {Seck}, \citenamefont {Shappert}, \citenamefont {Guise}, \citenamefont {Volin}, \citenamefont {Fallek} \emph {et~al.}}]{clark2021high}%
  \BibitemOpen
  \bibfield  {author} {\bibinfo {author} {\bibfnamefont {C.~R.}\ \bibnamefont {Clark}}, \bibinfo {author} {\bibfnamefont {H.~N.}\ \bibnamefont {Tinkey}}, \bibinfo {author} {\bibfnamefont {B.~C.}\ \bibnamefont {Sawyer}}, \bibinfo {author} {\bibfnamefont {A.~M.}\ \bibnamefont {Meier}}, \bibinfo {author} {\bibfnamefont {K.~A.}\ \bibnamefont {Burkhardt}}, \bibinfo {author} {\bibfnamefont {C.~M.}\ \bibnamefont {Seck}}, \bibinfo {author} {\bibfnamefont {C.~M.}\ \bibnamefont {Shappert}}, \bibinfo {author} {\bibfnamefont {N.~D.}\ \bibnamefont {Guise}}, \bibinfo {author} {\bibfnamefont {C.~E.}\ \bibnamefont {Volin}}, \bibinfo {author} {\bibfnamefont {S.~D.}\ \bibnamefont {Fallek}}, \emph {et~al.},\ }\bibfield  {title} {\bibinfo {title} {High-fidelity bell-state preparation with $^{40}\mathrm{Ca}^{+}$ optical qubits},\ }\href {https://doi.org/10.1103/PhysRevLett.127.130505} {\bibfield  {journal} {\bibinfo  {journal} {Phys. Rev. Lett.}\ }\textbf {\bibinfo {volume} {127}},\ \bibinfo {pages} {130505} (\bibinfo {year}
  {2021})}\BibitemShut {NoStop}%
\bibitem [{\citenamefont {L{\"o}schnauer}\ \emph {et~al.}(2024)\citenamefont {L{\"o}schnauer}, \citenamefont {Toba}, \citenamefont {Hughes}, \citenamefont {King}, \citenamefont {Weber}, \citenamefont {Srinivas}, \citenamefont {Matt}, \citenamefont {Nourshargh}, \citenamefont {Allcock}, \citenamefont {Ballance} \emph {et~al.}}]{loschnauer2024scalable}%
  \BibitemOpen
  \bibfield  {author} {\bibinfo {author} {\bibfnamefont {C.}~\bibnamefont {L{\"o}schnauer}}, \bibinfo {author} {\bibfnamefont {J.~M.}\ \bibnamefont {Toba}}, \bibinfo {author} {\bibfnamefont {A.}~\bibnamefont {Hughes}}, \bibinfo {author} {\bibfnamefont {S.}~\bibnamefont {King}}, \bibinfo {author} {\bibfnamefont {M.}~\bibnamefont {Weber}}, \bibinfo {author} {\bibfnamefont {R.}~\bibnamefont {Srinivas}}, \bibinfo {author} {\bibfnamefont {R.}~\bibnamefont {Matt}}, \bibinfo {author} {\bibfnamefont {R.}~\bibnamefont {Nourshargh}}, \bibinfo {author} {\bibfnamefont {D.}~\bibnamefont {Allcock}}, \bibinfo {author} {\bibfnamefont {C.}~\bibnamefont {Ballance}}, \emph {et~al.},\ }\bibfield  {title} {\bibinfo {title} {Scalable, high-fidelity all-electronic control of trapped-ion qubits},\ }\href {https://arxiv.org/abs/2407.07694} {\bibfield  {journal} {\bibinfo  {journal} {arXiv preprint arXiv:2407.07694}\ } (\bibinfo {year} {2024})}\BibitemShut {NoStop}%
\bibitem [{\citenamefont {Leibfried}\ \emph {et~al.}(2003)\citenamefont {Leibfried}, \citenamefont {Blatt}, \citenamefont {Monroe},\ and\ \citenamefont {Wineland}}]{leibfried2003quantum}%
  \BibitemOpen
  \bibfield  {author} {\bibinfo {author} {\bibfnamefont {D.}~\bibnamefont {Leibfried}}, \bibinfo {author} {\bibfnamefont {R.}~\bibnamefont {Blatt}}, \bibinfo {author} {\bibfnamefont {C.}~\bibnamefont {Monroe}},\ and\ \bibinfo {author} {\bibfnamefont {D.}~\bibnamefont {Wineland}},\ }\bibfield  {title} {\bibinfo {title} {Quantum dynamics of single trapped ions},\ }\href {https://doi.org/10.1103/RevModPhys.75.281} {\bibfield  {journal} {\bibinfo  {journal} {Rev. Mod. Phys.}\ }\textbf {\bibinfo {volume} {75}},\ \bibinfo {pages} {281} (\bibinfo {year} {2003})}\BibitemShut {NoStop}%
\bibitem [{\citenamefont {H{\"a}ffner}\ \emph {et~al.}(2008)\citenamefont {H{\"a}ffner}, \citenamefont {Roos},\ and\ \citenamefont {Blatt}}]{haffner2008quantum}%
  \BibitemOpen
  \bibfield  {author} {\bibinfo {author} {\bibfnamefont {H.}~\bibnamefont {H{\"a}ffner}}, \bibinfo {author} {\bibfnamefont {C.~F.}\ \bibnamefont {Roos}},\ and\ \bibinfo {author} {\bibfnamefont {R.}~\bibnamefont {Blatt}},\ }\bibfield  {title} {\bibinfo {title} {Quantum computing with trapped ions},\ }\href {https://doi.org/10.1016/j.physrep.2008.09.003} {\bibfield  {journal} {\bibinfo  {journal} {Phys. Rep.}\ }\textbf {\bibinfo {volume} {469}},\ \bibinfo {pages} {155} (\bibinfo {year} {2008})}\BibitemShut {NoStop}%
\bibitem [{\citenamefont {Chen}\ \emph {et~al.}(2021)\citenamefont {Chen}, \citenamefont {Gan}, \citenamefont {Zhang}, \citenamefont {Matuskevich},\ and\ \citenamefont {Kim}}]{chen2021quantum}%
  \BibitemOpen
  \bibfield  {author} {\bibinfo {author} {\bibfnamefont {W.}~\bibnamefont {Chen}}, \bibinfo {author} {\bibfnamefont {J.}~\bibnamefont {Gan}}, \bibinfo {author} {\bibfnamefont {J.-N.}\ \bibnamefont {Zhang}}, \bibinfo {author} {\bibfnamefont {D.}~\bibnamefont {Matuskevich}},\ and\ \bibinfo {author} {\bibfnamefont {K.}~\bibnamefont {Kim}},\ }\bibfield  {title} {\bibinfo {title} {Quantum computation and simulation with vibrational modes of trapped ions},\ }\href {https://doi.org/10.1088/1674-1056/ac01e3} {\bibfield  {journal} {\bibinfo  {journal} {Chin. Phys. B}\ }\textbf {\bibinfo {volume} {30}},\ \bibinfo {pages} {060311} (\bibinfo {year} {2021})}\BibitemShut {NoStop}%
\bibitem [{\citenamefont {Cai}\ \emph {et~al.}(2023)\citenamefont {Cai}, \citenamefont {Luan}, \citenamefont {Ou}, \citenamefont {Tu}, \citenamefont {Yin}, \citenamefont {Zhang},\ and\ \citenamefont {Kim}}]{cai2023entangling}%
  \BibitemOpen
  \bibfield  {author} {\bibinfo {author} {\bibfnamefont {Z.}~\bibnamefont {Cai}}, \bibinfo {author} {\bibfnamefont {C.-Y.}\ \bibnamefont {Luan}}, \bibinfo {author} {\bibfnamefont {L.}~\bibnamefont {Ou}}, \bibinfo {author} {\bibfnamefont {H.}~\bibnamefont {Tu}}, \bibinfo {author} {\bibfnamefont {Z.}~\bibnamefont {Yin}}, \bibinfo {author} {\bibfnamefont {J.-N.}\ \bibnamefont {Zhang}},\ and\ \bibinfo {author} {\bibfnamefont {K.}~\bibnamefont {Kim}},\ }\bibfield  {title} {\bibinfo {title} {Entangling gates for trapped-ion quantum computation and quantum simulation},\ }\href {https://doi.org/10.1007/s40042-023-00772-3} {\bibfield  {journal} {\bibinfo  {journal} {J. Korean Phys. Soc.}\ }\textbf {\bibinfo {volume} {82}},\ \bibinfo {pages} {882} (\bibinfo {year} {2023})}\BibitemShut {NoStop}%
\bibitem [{\citenamefont {H{\"a}ffner}\ \emph {et~al.}(2003)\citenamefont {H{\"a}ffner}, \citenamefont {Gulde}, \citenamefont {Riebe}, \citenamefont {Lancaster}, \citenamefont {Becher}, \citenamefont {Eschner}, \citenamefont {Schmidt-Kaler},\ and\ \citenamefont {Blatt}}]{haffner2003precision}%
  \BibitemOpen
  \bibfield  {author} {\bibinfo {author} {\bibfnamefont {H.}~\bibnamefont {H{\"a}ffner}}, \bibinfo {author} {\bibfnamefont {S.}~\bibnamefont {Gulde}}, \bibinfo {author} {\bibfnamefont {M.}~\bibnamefont {Riebe}}, \bibinfo {author} {\bibfnamefont {G.}~\bibnamefont {Lancaster}}, \bibinfo {author} {\bibfnamefont {C.}~\bibnamefont {Becher}}, \bibinfo {author} {\bibfnamefont {J.}~\bibnamefont {Eschner}}, \bibinfo {author} {\bibfnamefont {F.}~\bibnamefont {Schmidt-Kaler}},\ and\ \bibinfo {author} {\bibfnamefont {R.}~\bibnamefont {Blatt}},\ }\bibfield  {title} {\bibinfo {title} {Precision measurement and compensation of optical stark shifts for an ion-trap quantum processor},\ }\href {https://doi.org/10.1103/PhysRevLett.90.143602} {\bibfield  {journal} {\bibinfo  {journal} {Phys. Rev. Lett.}\ }\textbf {\bibinfo {volume} {90}},\ \bibinfo {pages} {143602} (\bibinfo {year} {2003})}\BibitemShut {NoStop}%
\bibitem [{\citenamefont {Blatt}\ and\ \citenamefont {Roos}(2012)}]{blatt2012quantum}%
  \BibitemOpen
  \bibfield  {author} {\bibinfo {author} {\bibfnamefont {R.}~\bibnamefont {Blatt}}\ and\ \bibinfo {author} {\bibfnamefont {C.~F.}\ \bibnamefont {Roos}},\ }\bibfield  {title} {\bibinfo {title} {Quantum simulations with trapped ions},\ }\href {https://doi.org/10.1038/nphys2252} {\bibfield  {journal} {\bibinfo  {journal} {Nat. Phys.}\ }\textbf {\bibinfo {volume} {8}},\ \bibinfo {pages} {277} (\bibinfo {year} {2012})}\BibitemShut {NoStop}%
\bibitem [{\citenamefont {Monroe}\ \emph {et~al.}(2021)\citenamefont {Monroe}, \citenamefont {Campbell}, \citenamefont {Duan}, \citenamefont {Gong}, \citenamefont {Gorshkov}, \citenamefont {Hess}, \citenamefont {Islam}, \citenamefont {Kim}, \citenamefont {Linke}, \citenamefont {Pagano} \emph {et~al.}}]{monroe2021programmable}%
  \BibitemOpen
  \bibfield  {author} {\bibinfo {author} {\bibfnamefont {C.}~\bibnamefont {Monroe}}, \bibinfo {author} {\bibfnamefont {W.~C.}\ \bibnamefont {Campbell}}, \bibinfo {author} {\bibfnamefont {L.-M.}\ \bibnamefont {Duan}}, \bibinfo {author} {\bibfnamefont {Z.-X.}\ \bibnamefont {Gong}}, \bibinfo {author} {\bibfnamefont {A.~V.}\ \bibnamefont {Gorshkov}}, \bibinfo {author} {\bibfnamefont {P.~W.}\ \bibnamefont {Hess}}, \bibinfo {author} {\bibfnamefont {R.}~\bibnamefont {Islam}}, \bibinfo {author} {\bibfnamefont {K.}~\bibnamefont {Kim}}, \bibinfo {author} {\bibfnamefont {N.~M.}\ \bibnamefont {Linke}}, \bibinfo {author} {\bibfnamefont {G.}~\bibnamefont {Pagano}}, \emph {et~al.},\ }\bibfield  {title} {\bibinfo {title} {Programmable quantum simulations of spin systems with trapped ions},\ }\href {https://doi.org/10.1103/RevModPhys.93.025001} {\bibfield  {journal} {\bibinfo  {journal} {Rev. Mod. Phys.}\ }\textbf {\bibinfo {volume} {93}},\ \bibinfo {pages} {025001} (\bibinfo {year} {2021})}\BibitemShut {NoStop}%
\bibitem [{\citenamefont {Gan}\ \emph {et~al.}(2018)\citenamefont {Gan}, \citenamefont {Maslennikov}, \citenamefont {Tseng}, \citenamefont {Tan}, \citenamefont {Kaewuam}, \citenamefont {Arnold}, \citenamefont {Matsukevich},\ and\ \citenamefont {Barrett}}]{gan2018oscillating}%
  \BibitemOpen
  \bibfield  {author} {\bibinfo {author} {\bibfnamefont {H.}~\bibnamefont {Gan}}, \bibinfo {author} {\bibfnamefont {G.}~\bibnamefont {Maslennikov}}, \bibinfo {author} {\bibfnamefont {K.-W.}\ \bibnamefont {Tseng}}, \bibinfo {author} {\bibfnamefont {T.}~\bibnamefont {Tan}}, \bibinfo {author} {\bibfnamefont {R.}~\bibnamefont {Kaewuam}}, \bibinfo {author} {\bibfnamefont {K.}~\bibnamefont {Arnold}}, \bibinfo {author} {\bibfnamefont {D.}~\bibnamefont {Matsukevich}},\ and\ \bibinfo {author} {\bibfnamefont {M.}~\bibnamefont {Barrett}},\ }\bibfield  {title} {\bibinfo {title} {Oscillating-magnetic-field effects in high-precision metrology},\ }\href {https://doi.org/10.1103/PhysRevA.98.032514} {\bibfield  {journal} {\bibinfo  {journal} {Phys. Rev. A}\ }\textbf {\bibinfo {volume} {98}},\ \bibinfo {pages} {032514} (\bibinfo {year} {2018})}\BibitemShut {NoStop}%
\bibitem [{\citenamefont {Ye}\ \emph {et~al.}(1999)\citenamefont {Ye}, \citenamefont {Vernooy},\ and\ \citenamefont {Kimble}}]{ye1999trapping}%
  \BibitemOpen
  \bibfield  {author} {\bibinfo {author} {\bibfnamefont {J.}~\bibnamefont {Ye}}, \bibinfo {author} {\bibfnamefont {D.}~\bibnamefont {Vernooy}},\ and\ \bibinfo {author} {\bibfnamefont {H.}~\bibnamefont {Kimble}},\ }\bibfield  {title} {\bibinfo {title} {Trapping of single atoms in cavity qed},\ }\href {https://doi.org/10.1103/PhysRevLett.83.4987} {\bibfield  {journal} {\bibinfo  {journal} {Phys. Rev. Lett.}\ }\textbf {\bibinfo {volume} {83}},\ \bibinfo {pages} {4987} (\bibinfo {year} {1999})}\BibitemShut {NoStop}%
\bibitem [{\citenamefont {Katori}\ \emph {et~al.}(1999)\citenamefont {Katori}, \citenamefont {Ido}, \citenamefont {Isoya},\ and\ \citenamefont {Kuwata-Gonokami}}]{katori1999magneto}%
  \BibitemOpen
  \bibfield  {author} {\bibinfo {author} {\bibfnamefont {H.}~\bibnamefont {Katori}}, \bibinfo {author} {\bibfnamefont {T.}~\bibnamefont {Ido}}, \bibinfo {author} {\bibfnamefont {Y.}~\bibnamefont {Isoya}},\ and\ \bibinfo {author} {\bibfnamefont {M.}~\bibnamefont {Kuwata-Gonokami}},\ }\bibfield  {title} {\bibinfo {title} {Magneto-optical trapping and cooling of strontium atoms down to the photon recoil temperature},\ }\href {https://doi.org/10.1103/PhysRevLett.82.1116} {\bibfield  {journal} {\bibinfo  {journal} {Phys. Rev. Lett.}\ }\textbf {\bibinfo {volume} {82}},\ \bibinfo {pages} {1116} (\bibinfo {year} {1999})}\BibitemShut {NoStop}%
\bibitem [{\citenamefont {Ido}\ and\ \citenamefont {Katori}(2003)}]{ido2003recoil}%
  \BibitemOpen
  \bibfield  {author} {\bibinfo {author} {\bibfnamefont {T.}~\bibnamefont {Ido}}\ and\ \bibinfo {author} {\bibfnamefont {H.}~\bibnamefont {Katori}},\ }\bibfield  {title} {\bibinfo {title} {Recoil-free spectroscopy of neutral sr atoms in the lamb-dicke regime},\ }\href {https://doi.org/PhysRevLett.91.053001} {\bibfield  {journal} {\bibinfo  {journal} {Phys. Rev. Lett.}\ }\textbf {\bibinfo {volume} {91}},\ \bibinfo {pages} {053001} (\bibinfo {year} {2003})}\BibitemShut {NoStop}%
\bibitem [{\citenamefont {McKeever}\ \emph {et~al.}(2003)\citenamefont {McKeever}, \citenamefont {Buck}, \citenamefont {Boozer}, \citenamefont {Kuzmich}, \citenamefont {N{\"a}gerl}, \citenamefont {Stamper-Kurn},\ and\ \citenamefont {Kimble}}]{mckeever2003state}%
  \BibitemOpen
  \bibfield  {author} {\bibinfo {author} {\bibfnamefont {J.}~\bibnamefont {McKeever}}, \bibinfo {author} {\bibfnamefont {J.}~\bibnamefont {Buck}}, \bibinfo {author} {\bibfnamefont {A.}~\bibnamefont {Boozer}}, \bibinfo {author} {\bibfnamefont {A.}~\bibnamefont {Kuzmich}}, \bibinfo {author} {\bibfnamefont {H.-C.}\ \bibnamefont {N{\"a}gerl}}, \bibinfo {author} {\bibfnamefont {D.}~\bibnamefont {Stamper-Kurn}},\ and\ \bibinfo {author} {\bibfnamefont {H.}~\bibnamefont {Kimble}},\ }\bibfield  {title} {\bibinfo {title} {State-insensitive cooling and trapping of single atoms in an optical cavity},\ }\href {https://doi.org/10.1103/PhysRevLett.90.133602} {\bibfield  {journal} {\bibinfo  {journal} {Phys. Rev. Lett.}\ }\textbf {\bibinfo {volume} {90}},\ \bibinfo {pages} {133602} (\bibinfo {year} {2003})}\BibitemShut {NoStop}%
\bibitem [{\citenamefont {Choi}\ and\ \citenamefont {Cho}(2007)}]{choi2007elimination}%
  \BibitemOpen
  \bibfield  {author} {\bibinfo {author} {\bibfnamefont {J.~M.}\ \bibnamefont {Choi}}\ and\ \bibinfo {author} {\bibfnamefont {D.}~\bibnamefont {Cho}},\ }\bibfield  {title} {\bibinfo {title} {Elimination of inhomogeneous broadening for a ground-state hyperfine transition in an optical trap},\ }\href {https://doi.org/10.1088/1742-6596/80/1/012037} {\bibfield  {journal} {\bibinfo  {journal} {J. Phys.: Conf. Ser.}\ }\textbf {\bibinfo {volume} {80}},\ \bibinfo {pages} {012037} (\bibinfo {year} {2007})}\BibitemShut {NoStop}%
\bibitem [{\citenamefont {Flambaum}\ \emph {et~al.}(2008)\citenamefont {Flambaum}, \citenamefont {Dzuba},\ and\ \citenamefont {Derevianko}}]{flambaum2008magic}%
  \BibitemOpen
  \bibfield  {author} {\bibinfo {author} {\bibfnamefont {V.}~\bibnamefont {Flambaum}}, \bibinfo {author} {\bibfnamefont {V.}~\bibnamefont {Dzuba}},\ and\ \bibinfo {author} {\bibfnamefont {A.}~\bibnamefont {Derevianko}},\ }\bibfield  {title} {\bibinfo {title} {Magic frequencies for cesium primary-frequency standard},\ }\href {https://doi.org/10.1103/PhysRevLett.101.220801} {\bibfield  {journal} {\bibinfo  {journal} {Phys. Rev. Lett.}\ }\textbf {\bibinfo {volume} {101}},\ \bibinfo {pages} {220801} (\bibinfo {year} {2008})}\BibitemShut {NoStop}%
\bibitem [{\citenamefont {Liu}\ \emph {et~al.}(2015)\citenamefont {Liu}, \citenamefont {Huang}, \citenamefont {Bian}, \citenamefont {Shao}, \citenamefont {Guan}, \citenamefont {Tang}, \citenamefont {Li}, \citenamefont {Mitroy},\ and\ \citenamefont {Gao}}]{liu2015measurement}%
  \BibitemOpen
  \bibfield  {author} {\bibinfo {author} {\bibfnamefont {P.-L.}\ \bibnamefont {Liu}}, \bibinfo {author} {\bibfnamefont {Y.}~\bibnamefont {Huang}}, \bibinfo {author} {\bibfnamefont {W.}~\bibnamefont {Bian}}, \bibinfo {author} {\bibfnamefont {H.}~\bibnamefont {Shao}}, \bibinfo {author} {\bibfnamefont {H.}~\bibnamefont {Guan}}, \bibinfo {author} {\bibfnamefont {Y.-B.}\ \bibnamefont {Tang}}, \bibinfo {author} {\bibfnamefont {C.-B.}\ \bibnamefont {Li}}, \bibinfo {author} {\bibfnamefont {J.}~\bibnamefont {Mitroy}},\ and\ \bibinfo {author} {\bibfnamefont {K.-L.}\ \bibnamefont {Gao}},\ }\bibfield  {title} {\bibinfo {title} {Measurement of magic wavelengths for the $^{40}\mathrm{Ca}^{+}$ clock transition},\ }\href {https://doi.org/10.1103/PhysRevLett.114.223001} {\bibfield  {journal} {\bibinfo  {journal} {Phys. Rev. Lett.}\ }\textbf {\bibinfo {volume} {114}},\ \bibinfo {pages} {223001} (\bibinfo {year} {2015})}\BibitemShut {NoStop}%
\bibitem [{\citenamefont {Brown}\ \emph {et~al.}(2017)\citenamefont {Brown}, \citenamefont {Phillips}, \citenamefont {Beloy}, \citenamefont {McGrew}, \citenamefont {Schioppo}, \citenamefont {Fasano}, \citenamefont {Milani}, \citenamefont {Zhang}, \citenamefont {Hinkley}, \citenamefont {Leopardi} \emph {et~al.}}]{brown2017hyperpolarizability}%
  \BibitemOpen
  \bibfield  {author} {\bibinfo {author} {\bibfnamefont {R.~C.}\ \bibnamefont {Brown}}, \bibinfo {author} {\bibfnamefont {N.~B.}\ \bibnamefont {Phillips}}, \bibinfo {author} {\bibfnamefont {K.}~\bibnamefont {Beloy}}, \bibinfo {author} {\bibfnamefont {W.~F.}\ \bibnamefont {McGrew}}, \bibinfo {author} {\bibfnamefont {M.}~\bibnamefont {Schioppo}}, \bibinfo {author} {\bibfnamefont {R.~J.}\ \bibnamefont {Fasano}}, \bibinfo {author} {\bibfnamefont {G.}~\bibnamefont {Milani}}, \bibinfo {author} {\bibfnamefont {X.}~\bibnamefont {Zhang}}, \bibinfo {author} {\bibfnamefont {N.}~\bibnamefont {Hinkley}}, \bibinfo {author} {\bibfnamefont {H.}~\bibnamefont {Leopardi}}, \emph {et~al.},\ }\bibfield  {title} {\bibinfo {title} {Hyperpolarizability and operational magic wavelength in an optical lattice clock},\ }\href {https://doi.org/10.1103/PhysRevLett.119.253001} {\bibfield  {journal} {\bibinfo  {journal} {Phys. Rev. Lett.}\ }\textbf {\bibinfo {volume} {119}},\ \bibinfo {pages} {253001} (\bibinfo {year} {2017})}\BibitemShut
  {NoStop}%
\bibitem [{\citenamefont {Rengelink}\ \emph {et~al.}(2018)\citenamefont {Rengelink}, \citenamefont {Van Der~Werf}, \citenamefont {Notermans}, \citenamefont {Jannin}, \citenamefont {Eikema}, \citenamefont {Hoogerland},\ and\ \citenamefont {Vassen}}]{rengelink2018precision}%
  \BibitemOpen
  \bibfield  {author} {\bibinfo {author} {\bibfnamefont {R.}~\bibnamefont {Rengelink}}, \bibinfo {author} {\bibfnamefont {Y.}~\bibnamefont {Van Der~Werf}}, \bibinfo {author} {\bibfnamefont {R.}~\bibnamefont {Notermans}}, \bibinfo {author} {\bibfnamefont {R.}~\bibnamefont {Jannin}}, \bibinfo {author} {\bibfnamefont {K.}~\bibnamefont {Eikema}}, \bibinfo {author} {\bibfnamefont {M.}~\bibnamefont {Hoogerland}},\ and\ \bibinfo {author} {\bibfnamefont {W.}~\bibnamefont {Vassen}},\ }\bibfield  {title} {\bibinfo {title} {Precision spectroscopy of helium in a magic wavelength optical dipole trap},\ }\href {https://doi.org/10.1038/s41567-018-0242-5} {\bibfield  {journal} {\bibinfo  {journal} {Nat. Phys.}\ }\textbf {\bibinfo {volume} {14}},\ \bibinfo {pages} {1132} (\bibinfo {year} {2018})}\BibitemShut {NoStop}%
\bibitem [{\citenamefont {Kim}\ \emph {et~al.}(2013)\citenamefont {Kim}, \citenamefont {Han},\ and\ \citenamefont {Cho}}]{kim2013magic}%
  \BibitemOpen
  \bibfield  {author} {\bibinfo {author} {\bibfnamefont {H.}~\bibnamefont {Kim}}, \bibinfo {author} {\bibfnamefont {H.~S.}\ \bibnamefont {Han}},\ and\ \bibinfo {author} {\bibfnamefont {D.}~\bibnamefont {Cho}},\ }\bibfield  {title} {\bibinfo {title} {Magic polarization for optical trapping of atoms without stark-induced dephasing},\ }\href {https://doi.org/10.1103/PhysRevLett.111.243004} {\bibfield  {journal} {\bibinfo  {journal} {Phys. Rev. Lett.}\ }\textbf {\bibinfo {volume} {111}},\ \bibinfo {pages} {243004} (\bibinfo {year} {2013})}\BibitemShut {NoStop}%
\bibitem [{\citenamefont {Jackson}\ and\ \citenamefont {Vutha}(2019)}]{jackson2019magic}%
  \BibitemOpen
  \bibfield  {author} {\bibinfo {author} {\bibfnamefont {S.}~\bibnamefont {Jackson}}\ and\ \bibinfo {author} {\bibfnamefont {A.~C.}\ \bibnamefont {Vutha}},\ }\bibfield  {title} {\bibinfo {title} {Magic polarization for cancellation of light shifts in two-photon optical clocks},\ }\href {https://doi.org/10.1103/PhysRevA.99.063422} {\bibfield  {journal} {\bibinfo  {journal} {Phys. Rev. A}\ }\textbf {\bibinfo {volume} {99}},\ \bibinfo {pages} {063422} (\bibinfo {year} {2019})}\BibitemShut {NoStop}%
\bibitem [{\citenamefont {Cho}(2023)}]{cho2023use}%
  \BibitemOpen
  \bibfield  {author} {\bibinfo {author} {\bibfnamefont {D.}~\bibnamefont {Cho}},\ }\bibfield  {title} {\bibinfo {title} {Use of vector polarizability to manipulate alkali-metal atoms},\ }\href {https://doi.org/10.1007/s40042-023-00776-z} {\bibfield  {journal} {\bibinfo  {journal} {J. Korean Phys. Soc.}\ }\textbf {\bibinfo {volume} {82}},\ \bibinfo {pages} {864} (\bibinfo {year} {2023})}\BibitemShut {NoStop}%
\bibitem [{\citenamefont {Vizvary}\ \emph {et~al.}(2024)\citenamefont {Vizvary}, \citenamefont {Wall}, \citenamefont {Boguslawski}, \citenamefont {Bareian}, \citenamefont {Derevianko}, \citenamefont {Campbell},\ and\ \citenamefont {Hudson}}]{vizvary2024eliminating}%
  \BibitemOpen
  \bibfield  {author} {\bibinfo {author} {\bibfnamefont {S.~R.}\ \bibnamefont {Vizvary}}, \bibinfo {author} {\bibfnamefont {Z.~J.}\ \bibnamefont {Wall}}, \bibinfo {author} {\bibfnamefont {M.~J.}\ \bibnamefont {Boguslawski}}, \bibinfo {author} {\bibfnamefont {M.}~\bibnamefont {Bareian}}, \bibinfo {author} {\bibfnamefont {A.}~\bibnamefont {Derevianko}}, \bibinfo {author} {\bibfnamefont {W.~C.}\ \bibnamefont {Campbell}},\ and\ \bibinfo {author} {\bibfnamefont {E.~R.}\ \bibnamefont {Hudson}},\ }\bibfield  {title} {\bibinfo {title} {Eliminating qubit-type cross-talk in the omg protocol},\ }\href {https://doi.org/10.1103/PhysRevLett.132.263201} {\bibfield  {journal} {\bibinfo  {journal} {Phys. Rev. Lett.}\ }\textbf {\bibinfo {volume} {132}},\ \bibinfo {pages} {263201} (\bibinfo {year} {2024})}\BibitemShut {NoStop}%
\bibitem [{\citenamefont {Mizrahi}\ \emph {et~al.}(2014)\citenamefont {Mizrahi}, \citenamefont {Neyenhuis}, \citenamefont {Johnson}, \citenamefont {Campbell}, \citenamefont {Senko}, \citenamefont {Hayes},\ and\ \citenamefont {Monroe}}]{mizrahi2014quantum}%
  \BibitemOpen
  \bibfield  {author} {\bibinfo {author} {\bibfnamefont {J.}~\bibnamefont {Mizrahi}}, \bibinfo {author} {\bibfnamefont {B.}~\bibnamefont {Neyenhuis}}, \bibinfo {author} {\bibfnamefont {K.}~\bibnamefont {Johnson}}, \bibinfo {author} {\bibfnamefont {W.}~\bibnamefont {Campbell}}, \bibinfo {author} {\bibfnamefont {C.}~\bibnamefont {Senko}}, \bibinfo {author} {\bibfnamefont {D.}~\bibnamefont {Hayes}},\ and\ \bibinfo {author} {\bibfnamefont {C.}~\bibnamefont {Monroe}},\ }\bibfield  {title} {\bibinfo {title} {Quantum control of qubits and atomic motion using ultrafast laser pulses},\ }\href {https://doi.org/10.1007/s00340-013-5717-6} {\bibfield  {journal} {\bibinfo  {journal} {Appl. Phys. B}\ }\textbf {\bibinfo {volume} {114}},\ \bibinfo {pages} {45} (\bibinfo {year} {2014})}\BibitemShut {NoStop}%
\bibitem [{\citenamefont {Hayes}\ \emph {et~al.}(2010)\citenamefont {Hayes}, \citenamefont {Matsukevich}, \citenamefont {Maunz}, \citenamefont {Hucul}, \citenamefont {Quraishi}, \citenamefont {Olmschenk}, \citenamefont {Campbell}, \citenamefont {Mizrahi}, \citenamefont {Senko},\ and\ \citenamefont {Monroe}}]{hayes2010entanglement}%
  \BibitemOpen
  \bibfield  {author} {\bibinfo {author} {\bibfnamefont {D.}~\bibnamefont {Hayes}}, \bibinfo {author} {\bibfnamefont {D.~N.}\ \bibnamefont {Matsukevich}}, \bibinfo {author} {\bibfnamefont {P.}~\bibnamefont {Maunz}}, \bibinfo {author} {\bibfnamefont {D.}~\bibnamefont {Hucul}}, \bibinfo {author} {\bibfnamefont {Q.}~\bibnamefont {Quraishi}}, \bibinfo {author} {\bibfnamefont {S.}~\bibnamefont {Olmschenk}}, \bibinfo {author} {\bibfnamefont {W.}~\bibnamefont {Campbell}}, \bibinfo {author} {\bibfnamefont {J.}~\bibnamefont {Mizrahi}}, \bibinfo {author} {\bibfnamefont {C.}~\bibnamefont {Senko}},\ and\ \bibinfo {author} {\bibfnamefont {C.}~\bibnamefont {Monroe}},\ }\bibfield  {title} {\bibinfo {title} {Entanglement of atomic qubits using an optical frequency comb},\ }\href {https://doi.org/10.1103/PhysRevLett.104.140501} {\bibfield  {journal} {\bibinfo  {journal} {Phys. Rev. Lett.}\ }\textbf {\bibinfo {volume} {104}},\ \bibinfo {pages} {140501} (\bibinfo {year} {2010})}\BibitemShut {NoStop}%
\bibitem [{\citenamefont {Campbell}\ \emph {et~al.}(2010)\citenamefont {Campbell}, \citenamefont {Mizrahi}, \citenamefont {Quraishi}, \citenamefont {Senko}, \citenamefont {Hayes}, \citenamefont {Hucul}, \citenamefont {Matsukevich}, \citenamefont {Maunz},\ and\ \citenamefont {Monroe}}]{campbell2010ultrafast}%
  \BibitemOpen
  \bibfield  {author} {\bibinfo {author} {\bibfnamefont {W.}~\bibnamefont {Campbell}}, \bibinfo {author} {\bibfnamefont {J.}~\bibnamefont {Mizrahi}}, \bibinfo {author} {\bibfnamefont {Q.}~\bibnamefont {Quraishi}}, \bibinfo {author} {\bibfnamefont {C.}~\bibnamefont {Senko}}, \bibinfo {author} {\bibfnamefont {D.}~\bibnamefont {Hayes}}, \bibinfo {author} {\bibfnamefont {D.}~\bibnamefont {Hucul}}, \bibinfo {author} {\bibfnamefont {D.}~\bibnamefont {Matsukevich}}, \bibinfo {author} {\bibfnamefont {P.}~\bibnamefont {Maunz}},\ and\ \bibinfo {author} {\bibfnamefont {C.}~\bibnamefont {Monroe}},\ }\bibfield  {title} {\bibinfo {title} {Ultrafast gates for single atomic qubits},\ }\href {https://doi.org/10.1103/PhysRevLett.105.090502} {\bibfield  {journal} {\bibinfo  {journal} {Phys. Rev. Lett.}\ }\textbf {\bibinfo {volume} {105}},\ \bibinfo {pages} {090502} (\bibinfo {year} {2010})}\BibitemShut {NoStop}%
\bibitem [{\citenamefont {Wong-Campos}\ \emph {et~al.}(2017)\citenamefont {Wong-Campos}, \citenamefont {Moses}, \citenamefont {Johnson},\ and\ \citenamefont {Monroe}}]{wong2017demonstration}%
  \BibitemOpen
  \bibfield  {author} {\bibinfo {author} {\bibfnamefont {J.~D.}\ \bibnamefont {Wong-Campos}}, \bibinfo {author} {\bibfnamefont {S.~A.}\ \bibnamefont {Moses}}, \bibinfo {author} {\bibfnamefont {K.~G.}\ \bibnamefont {Johnson}},\ and\ \bibinfo {author} {\bibfnamefont {C.}~\bibnamefont {Monroe}},\ }\bibfield  {title} {\bibinfo {title} {Demonstration of two-atom entanglement with ultrafast optical pulses},\ }\href {https://doi.org/10.1103/PhysRevLett.119.230501} {\bibfield  {journal} {\bibinfo  {journal} {Phys. Rev. Lett.}\ }\textbf {\bibinfo {volume} {119}},\ \bibinfo {pages} {230501} (\bibinfo {year} {2017})}\BibitemShut {NoStop}%
\bibitem [{\citenamefont {Lu}\ \emph {et~al.}(2019)\citenamefont {Lu}, \citenamefont {Zhang}, \citenamefont {Zhang}, \citenamefont {Chen}, \citenamefont {Shen}, \citenamefont {Zhang}, \citenamefont {Zhang},\ and\ \citenamefont {Kim}}]{lu2019global}%
  \BibitemOpen
  \bibfield  {author} {\bibinfo {author} {\bibfnamefont {Y.}~\bibnamefont {Lu}}, \bibinfo {author} {\bibfnamefont {S.}~\bibnamefont {Zhang}}, \bibinfo {author} {\bibfnamefont {K.}~\bibnamefont {Zhang}}, \bibinfo {author} {\bibfnamefont {W.}~\bibnamefont {Chen}}, \bibinfo {author} {\bibfnamefont {Y.}~\bibnamefont {Shen}}, \bibinfo {author} {\bibfnamefont {J.}~\bibnamefont {Zhang}}, \bibinfo {author} {\bibfnamefont {J.-N.}\ \bibnamefont {Zhang}},\ and\ \bibinfo {author} {\bibfnamefont {K.}~\bibnamefont {Kim}},\ }\bibfield  {title} {\bibinfo {title} {Global entangling gates on arbitrary ion qubits},\ }\href {https://doi.org/10.1038/s41586-019-1428-4} {\bibfield  {journal} {\bibinfo  {journal} {Nature (London)}\ }\textbf {\bibinfo {volume} {572}},\ \bibinfo {pages} {363} (\bibinfo {year} {2019})}\BibitemShut {NoStop}%
\bibitem [{\citenamefont {Figgatt}\ \emph {et~al.}(2019)\citenamefont {Figgatt}, \citenamefont {Ostrander}, \citenamefont {Linke}, \citenamefont {Landsman}, \citenamefont {Zhu}, \citenamefont {Maslov},\ and\ \citenamefont {Monroe}}]{figgatt2019parallel}%
  \BibitemOpen
  \bibfield  {author} {\bibinfo {author} {\bibfnamefont {C.}~\bibnamefont {Figgatt}}, \bibinfo {author} {\bibfnamefont {A.}~\bibnamefont {Ostrander}}, \bibinfo {author} {\bibfnamefont {N.~M.}\ \bibnamefont {Linke}}, \bibinfo {author} {\bibfnamefont {K.~A.}\ \bibnamefont {Landsman}}, \bibinfo {author} {\bibfnamefont {D.}~\bibnamefont {Zhu}}, \bibinfo {author} {\bibfnamefont {D.}~\bibnamefont {Maslov}},\ and\ \bibinfo {author} {\bibfnamefont {C.}~\bibnamefont {Monroe}},\ }\bibfield  {title} {\bibinfo {title} {Parallel entangling operations on a universal ion-trap quantum computer},\ }\href {https://doi.org/10.1038/s41586-019-1427-5} {\bibfield  {journal} {\bibinfo  {journal} {Nature (London)}\ }\textbf {\bibinfo {volume} {572}},\ \bibinfo {pages} {368} (\bibinfo {year} {2019})}\BibitemShut {NoStop}%
\bibitem [{\citenamefont {Lee}\ \emph {et~al.}(2016)\citenamefont {Lee}, \citenamefont {Smith}, \citenamefont {Richerme}, \citenamefont {Neyenhuis}, \citenamefont {Hess}, \citenamefont {Zhang},\ and\ \citenamefont {Monroe}}]{lee2016engineering}%
  \BibitemOpen
  \bibfield  {author} {\bibinfo {author} {\bibfnamefont {A.~C.}\ \bibnamefont {Lee}}, \bibinfo {author} {\bibfnamefont {J.}~\bibnamefont {Smith}}, \bibinfo {author} {\bibfnamefont {P.}~\bibnamefont {Richerme}}, \bibinfo {author} {\bibfnamefont {B.}~\bibnamefont {Neyenhuis}}, \bibinfo {author} {\bibfnamefont {P.~W.}\ \bibnamefont {Hess}}, \bibinfo {author} {\bibfnamefont {J.}~\bibnamefont {Zhang}},\ and\ \bibinfo {author} {\bibfnamefont {C.}~\bibnamefont {Monroe}},\ }\bibfield  {title} {\bibinfo {title} {Engineering large stark shifts for control of individual clock state qubits},\ }\href {https://doi.org/10.1103/PhysRevA.94.042308} {\bibfield  {journal} {\bibinfo  {journal} {Phys. Rev. A}\ }\textbf {\bibinfo {volume} {94}},\ \bibinfo {pages} {042308} (\bibinfo {year} {2016})}\BibitemShut {NoStop}%
\bibitem [{\citenamefont {Derevianko}(2010)}]{derevianko2010theory}%
  \BibitemOpen
  \bibfield  {author} {\bibinfo {author} {\bibfnamefont {A.}~\bibnamefont {Derevianko}},\ }\bibfield  {title} {\bibinfo {title} {Theory of magic optical traps for zeeman-insensitive clock transitions in alkali-metal atoms},\ }\href {https://doi.org/10.1103/PhysRevA.81.051606} {\bibfield  {journal} {\bibinfo  {journal} {Phys. Rev. A}\ }\textbf {\bibinfo {volume} {81}},\ \bibinfo {pages} {051606} (\bibinfo {year} {2010})}\BibitemShut {NoStop}%
\bibitem [{Sup()}]{Supp}%
  \BibitemOpen
  \href@noop {} {\bibinfo {title} {Supplementary materials}}\BibitemShut {NoStop}%
\bibitem [{\citenamefont {Wineland}\ \emph {et~al.}(2003)\citenamefont {Wineland}, \citenamefont {Barrett}, \citenamefont {Britton}, \citenamefont {Chiaverini}, \citenamefont {DeMarco}, \citenamefont {Itano}, \citenamefont {Jelenkovi{\'c}}, \citenamefont {Langer}, \citenamefont {Leibfried}, \citenamefont {Meyer} \emph {et~al.}}]{wineland2003quantum}%
  \BibitemOpen
  \bibfield  {author} {\bibinfo {author} {\bibfnamefont {D.~J.}\ \bibnamefont {Wineland}}, \bibinfo {author} {\bibfnamefont {M.}~\bibnamefont {Barrett}}, \bibinfo {author} {\bibfnamefont {J.}~\bibnamefont {Britton}}, \bibinfo {author} {\bibfnamefont {J.}~\bibnamefont {Chiaverini}}, \bibinfo {author} {\bibfnamefont {B.}~\bibnamefont {DeMarco}}, \bibinfo {author} {\bibfnamefont {W.~M.}\ \bibnamefont {Itano}}, \bibinfo {author} {\bibfnamefont {B.}~\bibnamefont {Jelenkovi{\'c}}}, \bibinfo {author} {\bibfnamefont {C.}~\bibnamefont {Langer}}, \bibinfo {author} {\bibfnamefont {D.}~\bibnamefont {Leibfried}}, \bibinfo {author} {\bibfnamefont {V.}~\bibnamefont {Meyer}}, \emph {et~al.},\ }\bibfield  {title} {\bibinfo {title} {Quantum information processing with trapped ions},\ }\href {https://doi.org/10.1098/rsta.2003.1205} {\bibfield  {journal} {\bibinfo  {journal} {Philos. Trans. R. Soc. A}\ }\textbf {\bibinfo {volume} {361}},\ \bibinfo {pages} {1349} (\bibinfo {year} {2003})}\BibitemShut {NoStop}%
\bibitem [{\citenamefont {Manakov}\ \emph {et~al.}(1986)\citenamefont {Manakov}, \citenamefont {Ovsiannikov},\ and\ \citenamefont {Rapoport}}]{manakov1986atoms}%
  \BibitemOpen
  \bibfield  {author} {\bibinfo {author} {\bibfnamefont {N.~L.}\ \bibnamefont {Manakov}}, \bibinfo {author} {\bibfnamefont {V.~D.}\ \bibnamefont {Ovsiannikov}},\ and\ \bibinfo {author} {\bibfnamefont {L.~P.}\ \bibnamefont {Rapoport}},\ }\bibfield  {title} {\bibinfo {title} {Atoms in a laser field},\ }\href {https://doi.org/10.1016/S0370-1573(86)80001-1} {\bibfield  {journal} {\bibinfo  {journal} {Phys. Rep.}\ }\textbf {\bibinfo {volume} {141}},\ \bibinfo {pages} {320} (\bibinfo {year} {1986})}\BibitemShut {NoStop}%
\bibitem [{\citenamefont {Le~Kien}\ \emph {et~al.}(2013)\citenamefont {Le~Kien}, \citenamefont {Schneeweiss},\ and\ \citenamefont {Rauschenbeutel}}]{le2013dynamical}%
  \BibitemOpen
  \bibfield  {author} {\bibinfo {author} {\bibfnamefont {F.}~\bibnamefont {Le~Kien}}, \bibinfo {author} {\bibfnamefont {P.}~\bibnamefont {Schneeweiss}},\ and\ \bibinfo {author} {\bibfnamefont {A.}~\bibnamefont {Rauschenbeutel}},\ }\bibfield  {title} {\bibinfo {title} {Dynamical polarizability of atoms in arbitrary light fields: general theory and application to cesium},\ }\href {https://doi.org/10.1140/epjd/e2013-30729-x} {\bibfield  {journal} {\bibinfo  {journal} {Eur. Phys. J. D}\ }\textbf {\bibinfo {volume} {67}},\ \bibinfo {pages} {92} (\bibinfo {year} {2013})}\BibitemShut {NoStop}%
\bibitem [{\citenamefont {Pinnington}\ \emph {et~al.}(1997)\citenamefont {Pinnington}, \citenamefont {Rieger},\ and\ \citenamefont {Kernahan}}]{pinnington1997beam}%
  \BibitemOpen
  \bibfield  {author} {\bibinfo {author} {\bibfnamefont {E.~H.}\ \bibnamefont {Pinnington}}, \bibinfo {author} {\bibfnamefont {G.}~\bibnamefont {Rieger}},\ and\ \bibinfo {author} {\bibfnamefont {J.~A.}\ \bibnamefont {Kernahan}},\ }\bibfield  {title} {\bibinfo {title} {Beam-laser measurements of the lifetimes of the $6p$ levels in $\mathrm{Yb}$ $\mathrm{II}$},\ }\href {https://doi.org/10.1103/PhysRevA.56.2421} {\bibfield  {journal} {\bibinfo  {journal} {Phys. Rev. A}\ }\textbf {\bibinfo {volume} {56}},\ \bibinfo {pages} {2421} (\bibinfo {year} {1997})}\BibitemShut {NoStop}%
\bibitem [{\citenamefont {Mizrahi}(2013)}]{mizrahi2013ultrafast}%
  \BibitemOpen
  \bibfield  {author} {\bibinfo {author} {\bibfnamefont {J.~A.}\ \bibnamefont {Mizrahi}},\ }\emph {\bibinfo {title} {Ultrafast control of spin and motion in trapped ions}},\ \href {https://www.proquest.com/dissertations-theses/ultrafast-control-spin-motion-trapped-ions/docview/1626728814/se-2} {Ph.D. thesis},\ \bibinfo  {school} {University of Maryland, College Park} (\bibinfo {year} {2013})\BibitemShut {NoStop}%
\end{thebibliography}%

\appendix
\clearpage
\onecolumngrid

\appendix
\counterwithin{equation}{section} 
\counterwithin{figure}{section}   
\counterwithin{table}{section}    


\section*{Supplementary materials for ``Precision Polarization Tuning for Light Shift Mitigation in Trapped-Ion Qubits"}
\setcounter{section}{0}
\setcounter{figure}{0}      
\setcounter{table}{0}      
\setcounter{equation}{0}   
\renewcommand{\thefigure}{\arabic{figure}}  
\renewcommand{\thetable}{\arabic{table}}    
\renewcommand{\theequation}{\arabic{equation}}  

\section{Supplementary Note 1 - Dressed State Derivation} 
\label{appendix:Dress State Derivation}

The formation of the dressed state arises primarily from the interaction between these two states, leading to the presence of off-diagonal elements in the matrix. Typically, since $g_J \ll g_I$ and $L = 0$ for the $^2S_{1/2}$ state, the interaction can be simplified as:
\begin{align}
    H_{\mathrm{hf}} =& w_{\mathrm{hf}} \mathbf{I} \cdot \mathbf{J} + 2\mu_B \mathbf{J} \cdot \mathbf{B} \notag \\
    =& w_{\mathrm{hf}} I_z J_z + \frac{w_{\mathrm{hf}}}{2}(I_+ J_- + I_-J_+) + 2\mu_B J_zB_z.
\label{equ:simplified hyperfine interaction}
\end{align}

Expanding the Hamiltonian in the four states $\ket{m_{I}, m_{J}}$, the result showed bellow, where $m_{I} = \pm \frac{1}{2}, m_{J} = \pm \frac{1}{2}, M = w_{\mathrm{hf}}, N = \mu_BB_z$.
\begin{equation}
\renewcommand{\arraystretch}{1.5} 
\begin{array}{c|cccc}
     (m_I,m_J)       & \ket{+\frac{1}{2}, +\frac{1}{2}} & \ket{+\frac{1}{2}, -\frac{1}{2}} & \ket{-\frac{1}{2}, +\frac{1}{2}} & \ket{-\frac{1}{2}, -\frac{1}{2}} \\
    \hline
    \bra{+\frac{1}{2}, +\frac{1}{2}} & \frac{1}{4}M + N & & &  \\
    \bra{+\frac{1}{2}, -\frac{1}{2}} & & -\frac{1}{4}M - N & \frac{1}{2}M &   \\
    \bra{-\frac{1}{2}, +\frac{1}{2}} & & \frac{1}{2}M & -\frac{1}{4}M + N &   \\
    \bra{-\frac{1}{2}, -\frac{1}{2}} & & & & \frac{1}{4}M - N \\
\end{array}
\label{equ:dress statematrix}
\end{equation}

In the absence of an external magnetic field, the four eigenstates of the matrix are
\begin{align}
    &\ket{F=0,m=0} = \frac{1}{\sqrt{2}}\left( \ket{-\frac{1}{2}, +\frac{1}{2}} -\ket{+\frac{1}{2}, -\frac{1}{2}} \right), \notag  \\
    &\ket{F=1,m=0} = \frac{1}{\sqrt{2}}\left( \ket{-\frac{1}{2}, +\frac{1}{2}} + \ket{+\frac{1}{2}, -\frac{1}{2}}\right),   \notag  \\
    &\ket{F=1,m=+1} = \ket{+\frac{1}{2}, +\frac{1}{2}}, \notag  \\
    &\ket{F=1,m=-1} = \ket{-\frac{1}{2}, -\frac{1}{2}},
\label{equ:dress state without B}
\end{align}

when involving the external magnetic field, the above eigenstates change to 
\begin{align}
    &\begin{aligned}
    \ket{F=0,m=0}_d &= A\left( \ket{-\frac{1}{2}, +\frac{1}{2}} - \left(2R+\sqrt{1+4R^2} \right)\ket{+\frac{1}{2}, -\frac{1}{2}} \right) \\ 
                    &= \ket{F=0,m=0} - R \ket{F=1,m=0} + o(R^2),
    \end{aligned}    \notag \\
    &\begin{aligned}
    \ket{F=1,m=0}_d &= A\left( \ket{-\frac{1}{2}, +\frac{1}{2}} - \left(2R-\sqrt{1+4R^2} \right)\ket{+\frac{1}{2}, -\frac{1}{2}} \right) \\
                   &= \ket{F=1,m=0} + R \ket{F=0,m=0} + o(R^2),
    &\end{aligned} \notag \\
    &\begin{aligned}
    \ket{F=1,m=+1}_d &= \ket{+\frac{1}{2}, +\frac{1}{2}} \\
                     &= \ket{F=1,m=+1},
    \end{aligned} \notag \\
    &\begin{aligned}
    \ket{F=1,m=-1}_d &= \ket{-\frac{1}{2}, -\frac{1}{2}} \\
                     &= \ket{F=1,m=-1},
    \end{aligned}
\label{equ:dress state with B}
\end{align}
which leads to the result of Eq.~\eqref{equ:dress state} and $ R = N/M $, $A$ is normalized coefficient.

\section{Supplementary Note 2 - Dipole Moment Perturbation of Dressed States}
\label{appendix:Dipole Moment Perturbation of Dressed States}
To obtain the dependence on the external magnetic field, the dipole moment, perturbed by the dressed state, needs to be calculated. As an example, we choose $\ket{0,0}_d \rightarrow \ket{^2P_{1/2},1,1}$ in the $\sigma_{+}$ polarization, which is given by

\begin{align}
    \bra{^2P_{1/2},1,1} \hat{\mathbf{d}} \ket{0,0}_d =& \bra{^2P_{1/2},1,1} \hat{\mathbf{d}} \ket{0,0} \notag \\
        - R& \bra{^2P_{1/2},1,1} \hat{\mathbf{d}} \ket{1,0} \notag \\
         =& \sqrt{\frac{2}{6}} d_1 + R \sqrt{\frac{2}{6}} d_1 = \sqrt{\frac{2}{6}}\left(1+R\right) d_1 .
\label{equ:calculate example}
\end{align}

Thus, the dipole moment changes from $\sqrt{\frac{2}{6}}$ to $\sqrt{\frac{2}{6}}\left(1+R\right)$. Similarly, the dipole moment for the two dressed states can also be revised, with the results summarized in Tab.~\ref{tab:CG coefficients}.

\begin{table}[ht]
\renewcommand{\arraystretch}{2.0}
    \centering
    \begin{tabular}{|c|c|c|c|} 
        \hline
        $\sigma_{\pm}$ & $\ket{^2P_{1/2},1,1}$ & $\ket{^2P_{3/2},1,1}$ & $\ket{^2P_{3/2},2,1}$ \\ 
        \hline
        $\ket{0,0}_d$ & $\sqrt{\frac{2}{6}}\left(1\pm R\right)$ & $\sqrt{\frac{4}{6}}\left(1\mp\frac{R}{2}\right)$ & $\mp\sqrt{\frac{3}{6}}R$ \\
        \hline
        $\ket{1,0}_d$ &$ -\sqrt{\frac{2}{6}}\left(1 \mp R\right)$ & $\sqrt{\frac{1}{6}}\left(1\pm2R\right)$ & $\sqrt{\frac{3}{6}}$ \\
        \hline
    \end{tabular}
    \caption{Dipole moment perturbed by the external magnetic field}
    \label{tab:CG coefficients}
\end{table}

\section{Supplementary Note 3 - Second-Order AC Stark Shift} 
\label{appendix:Second Order AC Stark Shift}
The AC Stark shift induced by the laser can be expressed as:
\begin{align}
    \Delta E_g^{(2)}  =- \sum_{e} \frac{\bra{g} \mathbf{\mathcal{E}}^* \cdot \mathbf{d}^* \ket{e}   \bra{e} \mathbf{d} \cdot \mathbf{\mathcal{E}} \ket{g} }{4(\omega_e-\omega_g - \omega_L)} \notag \\  -\sum_{e} \frac{\bra{e} \mathbf{\mathcal{E}}^* \cdot \mathbf{d}^* \ket{g}   \bra{g} \mathbf{d} \cdot \mathbf{\mathcal{E}} \ket{e} }{4(\omega_e-\omega_g + \omega_L)},
\label{equ:general ac shift}
\end{align}
where $\ket{e}$ is the state inducing the AC Stark shift to $\ket{g}$, $\mathbf{d}$ is the electronic dipole moment, $\mathbf{\mathcal{E}}$ is the laser electronic field, $\omega_e$, $\omega_g$ and $\omega_L$ is the energy of $\ket{e}$, $\ket{g}$ and laser.

When considering the coupling of the four states of the $^2S_{1/2}$ manifold to the excited state, the AC Stark shift of these four states can be calculated as follows. Under the condition that the laser is aligned along the magnetic field $\mathbf{B}$, the ion will only experience left- and right-circularly polarized light. This polarization constraint simplifies the calculation, as only these two components contribute to the Stark shift. The resulting AC Stark shifts for the four states can then be expressed: 
\begin{align}
\Delta E_{00}^{(2)} &= (\epsilon_+^2 + \epsilon_-^2)   \left(\frac{g_{1/2}^2I(t)}{12\Delta_{1/2}} + \frac{2g_{3/2}^2I(t)}{12\Delta_{3/2}} \right),  \notag &&\\
\Delta E_{10}^{(2)} &= (\epsilon_+^2 + \epsilon_-^2)   \left(\frac{g_{1/2}^2I(t)}{12(\Delta_{1/2}-\omega_{\mathrm{hf}})} + \frac{2g_{3/2}^2I(t)}{12(\Delta_{3/2}-\omega_{\mathrm{hf}})} \right),  \notag \\
\Delta E_{1+1}^{(2)} &= \epsilon_+^2 \left(\frac{3g_{3/2}^2I(t)}{12(\Delta_{3/2}-\omega_{\mathrm{hf}})} \right) + \epsilon_-^2  \left(\frac{2g_{1/2}^2I(t)}{12(\Delta_{1/2}-\omega_{\mathrm{hf}})} + \frac{g_{3/2}^2I(t)}{12(\Delta_{3/2}-\omega_{\mathrm{hf}})}\right), \notag   &&\\
\Delta E_{1-1}^{(2)} &=  \epsilon_+^2 \left(\frac{2g_{1/2}^2I(t)}{12(\Delta_{1/2}-\omega_{\mathrm{hf}})} + \frac{g_{3/2}^2I(t)}{12(\Delta_{3/2}-\omega_{\mathrm{hf}})}\right)  + \epsilon_-^2  \left(\frac{3g_{3/2}^2I(t)}{12(\Delta_{3/2}-\omega_{\mathrm{hf}})} \right)  ,
\label{equ:four states ac Stark shift}
\end{align}

where $g_{J}^2 = \frac{\gamma^2_{J}}{2\bar{I}_{J}}$,where $J=1/2$ or $3/2$, $\gamma$ is the spontaneous emission rate, $\Bar{I}$ is the saturation intensity, and $I(t)$ is the laser power. In the calculation, we explicitly neglect contributions from hyperfine splitting in the excited $^2P_{1/2}$ and $^2P_{3/2}$ states.

When using the above dipole moment to Eq.~\eqref{equ:four states ac Stark shift}, the AC Stark shifts of clock states are corrected to 

\begin{flalign}
\Delta E_{00}^{(2)} &=  \left[ \epsilon_+^2   \left(\frac{g_{1/2}^2}{12\Delta_{1/2}} \left( 1+R\right)^2 + \frac{2g_{3/2}^2} {12\Delta_{3/2}} \left( 1-\frac{R}{2}\right)^2 + \frac{3g_{3/2}^2} {24\Delta_{3/2}} R^2\right) \right. \notag \\
&\qquad\qquad\qquad\qquad\qquad + \left. \epsilon_-^2   \left(\frac{g_{1/2}^2}{12\Delta_{1/2}} \left( 1-R\right)^2 + \frac{2g_{3/2}^2} {12\Delta_{3/2}} \left( 1+\frac{R}{2}\right)^2 + \frac{3g_{3/2}^2} {24\Delta_{3/2}} R^2\right)  \right] I(t)\notag \\ 
&= \left[\left( \epsilon_+^2 +\epsilon_-^2 \right)\frac{1}{12} \left( \frac{g^2_{1/2}}{\Delta_{1/2}} + \frac{2g^2_{3/2}}{\Delta_{3/2}} \right)  + \left( \epsilon_+^2 -\epsilon_-^2 \right) \frac{2R}{12}  \left ( \frac{g^2_{1/2}}{\Delta_{1/2} } - \frac{g^2_{3/2}}{\Delta_{3/2}}  \right) + o(R^2) \right ] I(t),  \\
\Delta E_{10}^{(2)} &=  \left[ \epsilon_+^2   \left(\frac{g_{1/2}^2}{12 \left(\Delta_{1/2}-\omega_{\mathrm{hf}}\right)} \left( 1-R\right)^2 + \frac{g_{3/2}^2} {24\left(\Delta_{3/2}-\omega_{\mathrm{hf}}\right)} \left( 1+2R\right)^2 + \frac{3g_{3/2}^2} {24\left(\Delta_{3/2}-\omega_{\mathrm{hf}}\right)} \right) \right. \notag \\
&\qquad\qquad\qquad\qquad\qquad + \left. \epsilon_-^2   \left(\frac{g_{1/2}^2}{12 \left(\Delta_{1/2}-\omega_{\mathrm{hf}}\right)} \left( 1+R\right)^2 + \frac{g_{3/2}^2} {24\left(\Delta_{3/2}-\omega_{\mathrm{hf}}\right)} \left( 1-2R\right)^2 + \frac{3g_{3/2}^2} {24\left(\Delta_{3/2}-\omega_{\mathrm{hf}}\right)} \right)  \right] I(t)\notag \\ 
&= \left[\left( \epsilon_+^2 +\epsilon_-^2 \right)\frac{1}{12} \left( \frac{g^2_{1/2}}{\left(\Delta_{1/2}-\omega_{\mathrm{hf}}\right)} + \frac{2g^2_{3/2}}{\left(\Delta_{3/2}-\omega_{\mathrm{hf}}\right)} \right)  - \left( \epsilon_+^2 -\epsilon_-^2 \right) \frac{2R}{12}  \left ( \frac{g^2_{1/2}}{\left(\Delta_{1/2}-\omega_{\mathrm{hf}}\right)} - \frac{g^2_{3/2}}{\left(\Delta_{3/2}-\omega_{\mathrm{hf}}\right)}  \right) + o(R^2) \right ] I(t) .
\label{equ:revised clock qubit shift}
\end{flalign}
Considering the differential shift of the clock qubit and ignoring the higher orders, the result is:
\begin{flalign}
\delta E_{\text{10,00}}^{(2)} &\equiv \Delta E_{10}^{(2)} -  \Delta E_{00}^{(2)} \notag\\ 
&= \left[\left( \epsilon_+^2 +\epsilon_-^2 \right)\frac{\omega_{\mathrm{hf}}}{12} \left( \frac{g^2_{1/2}}{\Delta_{1/2}^2} + \frac{2g^2_{3/2}}{\Delta_{3/2}^2} \right)\right. \left.- \left( \epsilon_+^2 -\epsilon_-^2 \right) \frac{4R}{12}  \left ( \frac{g^2_{1/2}}{\Delta_{1/2} } - \frac{g^2_{3/2}}{\Delta_{3/2}}  \right) + o(R^2) \right] I(t) .
\end{flalign}

For the Zeeman qubits, the result is:
\begin{flalign}
&\delta E_{\text{1±1,00}}^{(2)} \equiv \Delta E_{1±1}^{(2)} -  \Delta E_{00}^{(2)} \notag\\
 &=  \left[ \epsilon_{\pm}^2   \left( -\frac{g_{1/2}^2}{12 \Delta_{1/2}} +\frac{g_{3/2}^2} {12\left(\Delta_{3/2}-\omega_{\mathrm{hf}}\right)}  + \frac{2\omega_{\mathrm{hf}}}{\Delta_{3/2}}\frac{g_{3/2}^2} {12\left(\Delta_{3/2}-\omega_{\mathrm{hf}}\right)}  \mp \frac{2R}{12} \left ( \frac{g^2_{1/2}}{\Delta_{1/2} } - \frac{g^2_{3/2}}{\Delta_{3/2}}  \right) \right) \right. \notag \\
&+ \left. \epsilon_\mp^2   \left(\frac{g_{1/2}^2}{12 \left(\Delta_{1/2}-\omega_{\mathrm{hf}}\right)}  - \frac{g_{3/2}^2} {12\left(\Delta_{3/2}-\omega_{\mathrm{hf}}\right)}  + \frac{2\omega_{\mathrm{hf}}}{\Delta_{3/2}}\frac{g_{3/2}^2} {12\left(\Delta_{3/2}-\omega_{\mathrm{hf}}\right)} + \frac{\omega_{\mathrm{hf}}}{\Delta_{1/2}} \frac{g_{1/2}^2}{12 \left(\Delta_{1/2}-\omega_{\mathrm{hf}}\right)}\pm\frac{2R}{12} \left ( \frac{g^2_{1/2}}{\Delta_{1/2} } - \frac{g^2_{3/2}}{\Delta_{3/2}}  \right)\right)  \right] I(t) \notag \\ 
&=\pm\left( \epsilon_-^2 -\epsilon_+^2 \right)\frac{1}{12}\left ( \frac{g^2_{1/2}}{\Delta_{1/2} } -  \frac{g^2_{3/2}}{\Delta_{3/2}}  \right)I(t) +  o(R) + o(\frac{\omega_{\mathrm{hf}}}{\Delta_{J}}).
\label{equ:revised Zeeman qubits ac Stark shift}
\end{flalign} 

In the main text, we neglect the small value terms in Zeeman qubits associated with the factor $R$ and the ratio $\frac{\omega_{\mathrm{hf}}}{\Delta_{J}}$, where $J = 1/2$ or $1/3$. These contributions are approximately three orders of magnitude smaller than the primary energy shift and thus have a negligible effect.

\section{Supplementary Note 4 - Numerical Estimation of AC Stark Shift}
\label{appendix:numerical ac Stark shift}
To get the concrete number of the shift, the numerical values are calculated from physical values in Tab.~\ref{tab:The value of physical parameter}.
\begin{table}[htbp]
\renewcommand{\arraystretch}{2.0}
    \centering
    \begin{tabular}{|c|c|c|c|} 
        \hline
        $\omega_{\mathrm{hf}}$ & 12.624812 GHz & $\mathbf{\mathcal{C}}_{10,00} $ & 0.971\\ 
        \hline
        $\gamma_{1/2}$ & 19.703 MHz ~\cite{pinnington1997beam} & $\gamma_{3/2}$  &  25.895 MHz~\cite{pinnington1997beam}\\
        \hline
        $\Bar{I}_{1/2}$        & 510.3 $W/m^2$ ~\cite{mizrahi2013ultrafast} & $\Bar{I}_{3/2}$  &  950.6 $W/m^2$ ~\cite{mizrahi2013ultrafast}\\
        \hline
        $g_{1/2}^2$ & 0.380 $\times 10^{12}$ s/kg & $g_{3/2}^2$  &  0.353 $\times 10^{12}$ s/kg\\
        \hline
         $\Delta_{1/2}$&  $+(2\pi)~33$ THz & $\Delta_{3/2}$& $-(2\pi)~67$ THz  \\
        \hline
    \end{tabular}
    \caption{The value of physical parameter for the $^{171}\text{Yb}^{+}$ ion}
    \label{tab:The value of physical parameter}
\end{table}

Thus, the final AC Stark shift can be calculated as
\begin{itemize}
\item For the clock qubit $\ket{m=0}$:
\begin{align}
\delta E_{10,00} &= \Delta E_{1,0} - \Delta E_{0,0} =  \delta E_{\mathrm{hf}}^{(2)} + \delta E_{\mathrm{hf}}^{(4)} \notag \\
               &=\left( \epsilon_+^2 +\epsilon_-^2 \right)5.337\times 10^{-7} \frac{I(t)}{W/m^2} \mathrm{Hz}    \notag \\
               &-\left( \epsilon_+^2 -\epsilon_-^2 \right) 7.066 \times 10^{-6} \frac{I(t)}{W/m^2} \mathrm{Hz}   \notag \\
               &+\left( \epsilon_+^2 -\epsilon_-^2 \right)^2 1.327 \times 10^{-13}\left( \frac{I(t)}{W/m^2}\right) ^2  \mathrm{Hz},
\label{equ:numerial ac shif 1}
\end{align}

\item For the Zeeman qubit $\ket{m=+1}$:

\begin{align}
\delta E_{1+1,00} &= \Delta E_{1+1} - \Delta E_{00} \notag \\
               &=-\left( \epsilon_+^2 -\epsilon_-^2 \right)1.40\times 10^{-3} \frac{I(t)}{W/m^2} \mathrm{Hz},  
\label{equ:numerial ac shif 2}
\end{align}

\item For the Zeeman qubit $\ket{m=-1}$:

\begin{align}
\delta E_{1-1,00} &= \Delta E_{1-1} - \Delta E_{00} \notag \\
               &= \left( \epsilon_+^2 -\epsilon_-^2 \right)1.40 \times 10^{-3} \frac{I(t)}{W/m^2} \mathrm{Hz}. 
\label{equ:numerial ac shif 3}
\end{align}
\end{itemize}

\section{Supplementary Note 5 - The Relation between the Polarization and Relative Wave Plate Angle \texorpdfstring{$\theta$}{theta}}
The $\sigma_+$ and $\sigma_-$ circular polarization can be expressed as:
\begin{equation}
E_{\sigma_+} = \frac{1}{\sqrt{2}}
\begin{bmatrix}
1 \\
-i \\
\end{bmatrix},
\end{equation}

\begin{equation}
E_{\sigma_-} = \frac{1}{\sqrt{2}}
\begin{bmatrix}
1 \\
i \\
\end{bmatrix}.
\end{equation}

The Jones matrix of the wave-plate can be expressed as:
\begin{equation}
J =
\begin{bmatrix}
\cos^2{\theta} + \sin^2{\theta} e^{i\delta} & \sin{\theta} \cos{\theta}(1 - e^{i\delta}) \\
\sin{\theta} \cos{\theta}(1 - e^{i\delta}) & \sin^2{\theta} + \cos^2{\theta} e^{i\delta} \\
\end{bmatrix},
\label{equ:jones matrix}
\end{equation}

where $\delta$ represents the phase delay between the fast and slow axes of the wave-plate, and $\theta$ is the angle between the fast axis and the horizontal $x$-axis. In this case, the QWP has a matrix corresponding to $\phi = \pi/2$, and the HWP corresponds to $\phi = \pi$.

In the experiment, the polarization is altered by rotating the HWP. Assuming the QWP is fixed with $\theta = 0$, the polarization changes as the angle of the HWP, $\theta$, is varied. Starting with an initial horizontal polarization, the polarization after rotation can be expressed as:

\begin{equation}
\begin{bmatrix}
1 & 0 \\
0 & i
\end{bmatrix}_{\substack{\text{QWP}}}
\cdot
\begin{bmatrix}
\cos2\theta & \sin2\theta  \\
\sin2\theta & -\cos2\theta
\end{bmatrix}_{\substack{\text{HWP}}}
\cdot
\begin{bmatrix}
1 \\
0
\end{bmatrix}_{\substack{\text{initial}}}
= \begin{bmatrix}
\cos2\theta \\
i\sin2\theta
\end{bmatrix}.
\end{equation}

Therefore, the coefficient can be expressed as:

\begin{equation}
\epsilon_+ = \frac{1}{\sqrt{2}} 
\begin{bmatrix}
1,i
\end{bmatrix}
\cdot
 \begin{bmatrix}
\cos2\theta \\
i\sin2\theta
\end{bmatrix} = \cos(2\theta + \pi/4),
\end{equation}

\begin{equation}
\epsilon_- = \frac{1}{\sqrt{2}} 
\begin{bmatrix}
1,-i
\end{bmatrix}
\cdot
 \begin{bmatrix}
\cos2\theta \\
i\sin2\theta
\end{bmatrix} = \sin(2\theta + \pi/4).
\end{equation}

\section{Supplementary Note 6 - Polarization of Minimum Differential Shift for Clock Qubit}
\label{appendix:Polarization of Minimum Differential Shift for Clock Qubit}
\begin{figure}[htbp]
    \centering
    \captionsetup{justification=raggedright}
    \includegraphics[width=0.95\textwidth]{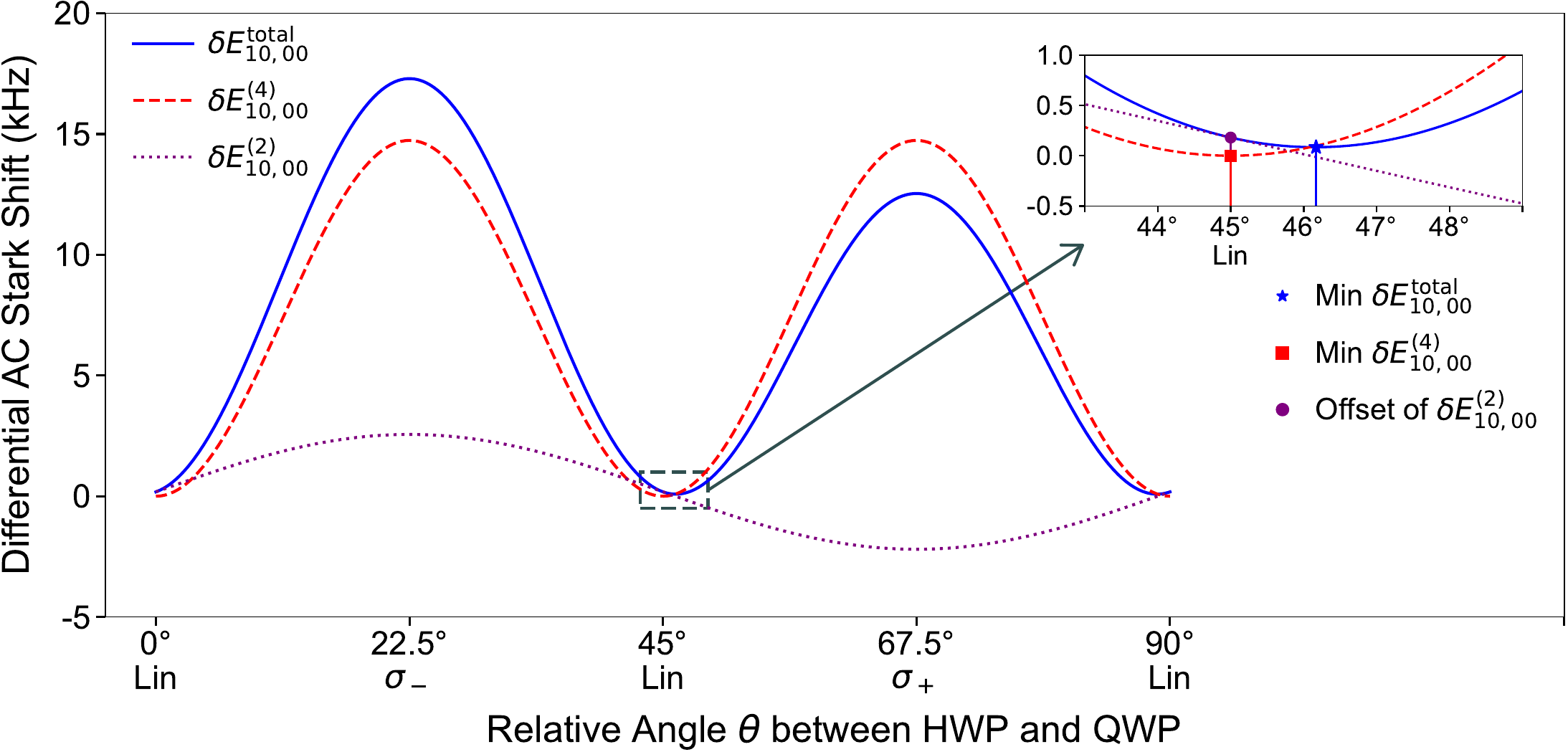}  
    \caption{\justifying Polarization-dependent differential shift of the clock qubit. The asymmetry induced by the external magnetic field shifts the position of the overall minimum (blue star) away from the fourth-order minimum (red square). The purple-filled circle indicates the offset of the second-order AC Stark shift in the absence of the magnetic field. The inset highlights the displacement of the minimal point due to magnetic-field-induced asymmetry.}
    \label{fig:polarization of minimal ac Stark shift}
\end{figure}

In this sequence, we illustrate how the minimum AC Stark shift of the clock qubit is affected by the presence of an external magnetic field. The curves in the figure represent simulated AC Stark shifts under a laser power of 52.1 mW and a beam waist radius of 7~$\mathrm{\mu m}$. As shown in Fig.~\ref{fig:polarization of minimal ac Stark shift}, the blue star marks the overall minimum of the total AC Stark shift, while the red square indicates the minimum of the fourth-order contribution. The purple-filled circle represents the offset of the second-order AC Stark shift, corresponding to the shift in the absence of an external magnetic field($B=0$) as defined in Eq.~\eqref{equ:clock qubit 2nd order}.
Due to the magnetic field, the AC Stark shift exhibits asymmetric behavior for polarization, causing a displacement of the overall minimum. As illustrated in the inset, this shifts the optimal point from the red square to the blue star. 

\begin{figure}[htbp]
    \centering
    \captionsetup{justification=raggedright}
    \includegraphics[width=0.95\textwidth]{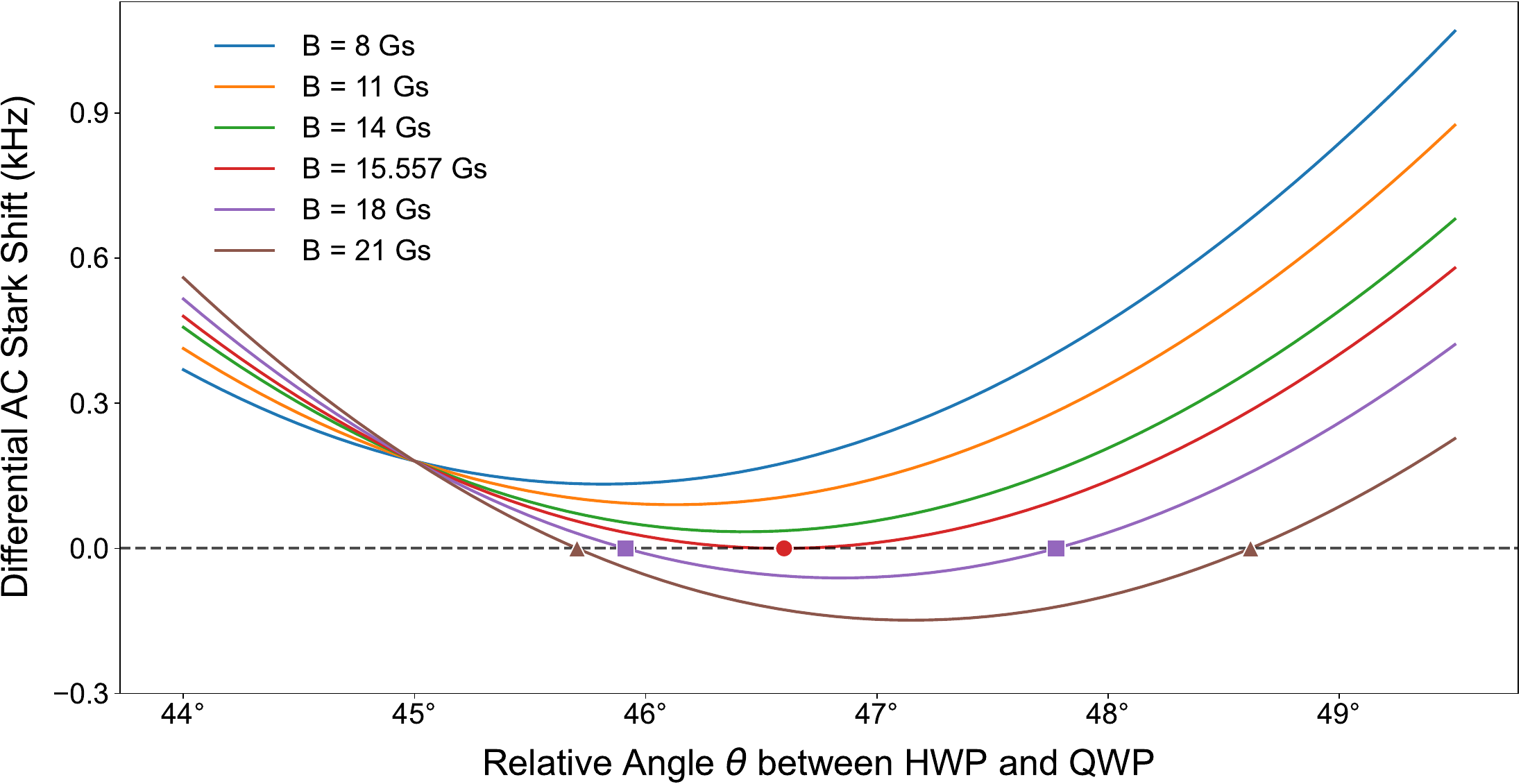}  
    \caption{\justifying Calculated AC Stark shift as a function of the relative angle under different magnetic field strengths. The calculation assumes a wavelength of 355 nm, an optical power of 52.1~mW, and a beam waist of 7~$\mathrm{\mu m}$. The solid lines represent the theoretical shifts for various magnetic fields, while the filled circles, squares, and triangles mark the corresponding zero-crossing points. No zero-crossing is observed when the magnetic field is below 15.557~Gauss, indicating that the light shift cannot be fully canceled. In contrast, for fields exceeding 15.557~Gauss, zero-crossing points emerge, allowing for perfect cancellation of the AC Stark shift via magic polarization.}
    \label{fig:theoritical magnetic field to cancel shift}
\end{figure}

Our simulations indicate that residual shifts persist under the current experimental conditions. As shown in Fig.~\ref{fig:theoritical magnetic field to cancel shift}, the magnitude of these shifts is related to the applied magnetic field strength. When the field strength exceeds 15.557~Gauss, the shift function crosses zero. By tuning the polarization to operate at these zero-crossing points, the AC Stark shift can be completely canceled, a condition known as magic polarization.


\end{document}